%% file: main.tex
\documentclass[sigplan,nonacm,screen]{acmart}
\usepackage{amsmath}
\usepackage[ruled,vlined,linesnumbered]{algorithm2e}
\usepackage{booktabs}
\usepackage{enumitem}
\usepackage{listings}
\usepackage{microtype}
\usepackage{tabularx}
\usepackage{tikz}
\usepackage{xspace}
\usetikzlibrary{arrows.meta,backgrounds,calc,decorations.pathreplacing,
  fit,matrix,positioning,shapes.geometric}

\definecolor{paperblue}{HTML}{2E75B6}
\definecolor{paperteal}{HTML}{4F8E78}
\definecolor{paperorange}{HTML}{ED7D31}
\definecolor{paperred}{HTML}{BC1E23}
\definecolor{paperpurple}{HTML}{7030A0}
\definecolor{paperink}{HTML}{2E3540}
\definecolor{papergray}{HTML}{F7FAFC}
\definecolor{paperlightblue}{HTML}{DEEBF7}
\definecolor{paperlightteal}{HTML}{E4EDD2}
\definecolor{paperlightorange}{HTML}{FBE5D6}
\definecolor{paperalg}{HTML}{4F8E78}
\definecolor{paperstate}{HTML}{2E75B6}
\definecolor{paperphys}{HTML}{ED7D31}
\definecolor{paperkernel}{HTML}{1A345F}
\definecolor{paperline}{HTML}{687683}
\definecolor{paperlightalg}{HTML}{E4EDD2}
\definecolor{paperlightstate}{HTML}{DEEBF7}
\definecolor{paperlightphys}{HTML}{FBE5D6}
\definecolor{paperlightkernel}{HTML}{D9E2F3}

\input{figures/style}

\newcommand{\ir}[1]{\textsf{#1}}
\newcommand{\code}[1]{\texttt{#1}}
\newcommand{\system}{\textsc{EquiForge}\xspace}

\input{figures/performance-statistics}

\newcolumntype{L}[1]{>{\raggedright\arraybackslash}p{#1}}
\newcolumntype{Y}{>{\raggedright\arraybackslash}X}

\lstdefinelanguage{EquiForgeIR}{
  sensitive=true,
  alsoletter={_},
  morecomment=[l]{//},
  morekeywords={func,where,return,eclass},
  morekeywords=[2]{input,tuple,get,partition,replicate,combination,collection,
    tile_matmul,matmul,reduce,fill,cast,multiply,mul,divide,subtract,add,
    maximum,minimum,max,min,exp,sqrt,rsqrt,transpose},
  morekeywords=[3]{Tensor,tensor,bf16,f16,f32,i32},
  morekeywords=[4]{axis,parallel,reducer,field,acc,to,permutation,keep_dims},
}

\lstdefinestyle{irsnippet}{
  language=EquiForgeIR,
  basicstyle=\normalsize\ttfamily\color{black},
  keywordstyle=\bfseries\color{black},
  keywordstyle=[2]\bfseries\color{paperblue!78!black},
  keywordstyle=[3]\color{paperpurple},
  keywordstyle=[4]\color{paperteal!92!black},
  commentstyle=\color{paperline!82},
  backgroundcolor=\color{white},
  frame=tb,
  framerule=0.35pt,
  rulecolor=\color{paperline!48},
  framesep=3pt,
  numbers=none,
  columns=fullflexible,
  keepspaces=true,
  showstringspaces=false,
  breaklines=false,
  escapeinside={(*@}{@*)},
}

\lstdefinestyle{irpaperexample}{
  style=irsnippet,
  frame=none,
  xleftmargin=0pt,
  xrightmargin=0pt,
  emph={partition,replicate,combination,collection,reduce},
  emphstyle=\bfseries\color{paperteal!92!black},
}

\lstdefinestyle{irteaser}{
  language=EquiForgeIR,
  basicstyle=\footnotesize\ttfamily\color{black},
  keywordstyle=\bfseries\color{black},
  keywordstyle=[2]\bfseries\color{paperblue!78!black},
  keywordstyle=[3]\color{paperpurple},
  keywordstyle=[4]\color{paperteal!92!black},
  commentstyle=\color{paperline!82},
  backgroundcolor=\color{white},
  frame=tb,
  framerule=0.35pt,
  rulecolor=\color{paperline!48},
  numbers=none,
  xleftmargin=0.15em,
  xrightmargin=0.15em,
  framesep=2.5pt,
  aboveskip=0.35em,
  belowskip=0.25em,
  columns=fullflexible,
  keepspaces=true,
  showstringspaces=false,
  breaklines=false,
  escapeinside={(*@}{@*)},
}

\title{Unleashing the Power of Equality Saturation for Tensor Program
Superoptimization}

\author{Qi Zhan}
\orcid{0000-0002-6800-1857}
\affiliation{%
  \institution{Zhejiang University}
  \city{Hangzhou}
  \country{China}
}
\email{qizhan@zju.edu.cn}

\author{Xing Hu}
\authornote{Corresponding author.}
\orcid{0000-0003-0093-3292}
\affiliation{%
  \institution{Zhejiang University}
  \city{Hangzhou}
  \country{China}
}
\email{xinghu@zju.edu.cn}

\author{Xin Xia}
\orcid{0000-0002-6302-3256}
\affiliation{%
  \institution{Zhejiang University}
  \city{Hangzhou}
  \country{China}
}
\affiliation{%
  \institution{Hangzhou High-Tech Zone (Binjiang) Institute of Blockchain and Data Security}
  \city{Hangzhou}
  \country{China}
}
\email{xin.xia@acm.org}

\author{Shanping Li}
\orcid{0000-0003-2615-9792}
\affiliation{%
  \institution{Zhejiang University}
  \city{Hangzhou}
  \country{China}
}
\email{shan@zju.edu.cn}

\renewcommand{\shortauthors}{Zhan et al.}
\hypersetup{pdfauthor={Qi Zhan, Xing Hu, Xin Xia, Shanping Li}}

\begin{document}

\input{sections/00-abstract}

\maketitle

\input{sections/01-introduction}
\input{sections/02-equality-saturation}
\input{sections/03-unified-ir}
\input{sections/04-equality-rules}
\input{sections/05-scalable-program-extraction}
\input{sections/06-code-generation-and-tuning}
\input{sections/07-evaluation}
\input{sections/08-related-work}
\input{sections/09-conclusion}

\bibliographystyle{ACM-Reference-Format}
\bibliography{references}

\end{document}

%% file: figures/style.tex
\tikzset{
  ef arrow/.style={-{Latex[length=1.25mm,width=0.85mm]},
    draw=paperline, line width=0.48pt},
  ef line/.style={draw=paperline, line width=0.48pt},
  ef guide/.style={draw=paperline!38, line width=0.35pt},
  ef tensor/.style={draw=paperalg!85!black, fill=paperalg,
    text=white, line width=0.45pt, minimum width=4.3mm,
    minimum height=4.3mm, inner sep=1pt, font=\sffamily},
  ef op/.style={draw=paperalg!85!black, fill=paperlightalg,
    rounded corners=2.4mm, line width=0.45pt, minimum height=4.5mm,
    inner xsep=3.0pt, inner ysep=1.4pt, font=\sffamily,
    align=center},
  ef state/.style={draw=paperstate, fill=white, line width=0.65pt,
    circle, minimum size=5.0mm, inner sep=0.6pt, font=\sffamily,
    align=center},
  ef state op/.style={draw=paperstate, fill=paperlightstate,
    rounded corners=2.4mm, line width=0.55pt, minimum height=4.8mm,
    inner xsep=3pt, inner ysep=1.3pt, font=\sffamily,
    align=center},
  ef tile/.style={draw=paperphys, fill=paperlightphys,
    rounded corners=0.8pt, line width=0.45pt, minimum height=5.2mm,
    inner xsep=2.5pt, inner ysep=1.2pt, font=\sffamily,
    align=center},
  ef memory/.style={draw=paperphys, fill=paperlightphys,
    rounded corners=0.7pt, line width=0.55pt, minimum height=4.5mm,
    inner xsep=3pt, inner ysep=1.2pt, font=\sffamily,
    align=center},
  ef kernel/.style={draw=paperkernel, fill=paperlightkernel,
    rounded corners=0.8pt, line width=0.55pt, minimum height=5.0mm,
    inner xsep=3pt, inner ysep=1.2pt, font=\sffamily,
    align=center},
  ef eclass/.style={draw=paperstate, dashed, dash pattern=on 2pt off 1.5pt,
    rounded corners=5mm, line width=0.5pt, fill=white},
  ef projection/.style={draw=paperstate, fill=paperstate, text=white,
    rounded corners=2.2mm, line width=0.45pt, minimum height=4.5mm,
    inner xsep=3.2pt, inner ysep=1.2pt, font=\sffamily},
  ef panel title/.style={anchor=west, font=\sffamily\bfseries,
    text=paperink},
  ef note/.style={font=\sffamily, text=paperline, align=center},
}

%% file: figures/performance-statistics.tex
\newcommand{\ProgramCases}{21}
\newcommand{\ProgramWins}{18}
\newcommand{\ProgramSpeedup}{1.32}
\newcommand{\ProgramMaxSpeedup}{2.74}

\newcommand{\MHADecodeMaxSpeedup}{1.87}
\newcommand{\MHAPrefillGapMin}{3}
\newcommand{\MHAPrefillGapMax}{6}

\newcommand{\LayerNormGapMin}{6}
\newcommand{\LayerNormGapMax}{12}
\newcommand{\PrismBEightCompileSpeedup}{2.74}

%% file: sections/00-abstract.tex
\begin{abstract}
Efficient GPU implementations of tensor programs often require
joint optimization of high-level algebraic formulations and
low-level execution strategies. However, the resulting search space grows rapidly as
transformations combine across operators, making joint
optimization difficult to scale. We present \system, a tensor
program superoptimizer based on equality saturation.
Its unified IR represents high-level tensor expressions and
tiled computations in a single expression language.
By composing equality rules, \system derives fused implementations
such as FlashAttention-style kernels directly from tensor expressions.
Early compaction prunes redundant
partial programs before completion, while subgraph composition
extends the search to larger graphs.
Across tensor-program benchmarks, \system achieves a geometric mean
speedup of $\ProgramSpeedup\times$ and a maximum of
$\ProgramMaxSpeedup\times$ over the fastest available baseline per
configuration. Its attention kernels outperform FlashAttention by
up to $\MHADecodeMaxSpeedup\times$ in decode and approach its performance in prefill.
\system also discovers new implementations that outperform
\texttt{torch.compile} on various Transformer layers, including
QK-normalized MLA ($3.16\times$) and mHC ($5.84\times$).
\end{abstract}

%% file: sections/01-introduction.tex
\section{Introduction}
\label{sec:introduction}

Efficient GPU execution is essential to the training and inference
of large language models. Tensor compilers translate high-level
tensor computations into efficient GPU programs through operator
fusion, tiling, and parallelization~\cite{tvm,tensorir,pytorch2}.
The same tensor computation can have many equivalent implementations
with substantially different performance.

\input{figures/overview}

Achieving high performance often requires rethinking both a computation's
mathematical formulation and its execution strategy.  For example, a key idea
in FlashAttention~\cite{flashattention,flashattention2} is to combine an online
reformulation of softmax with tiled implementations of attention's two matrix
multiplications.  Together, these transformations allow attention to be computed
incrementally within a fused kernel, without storing the full score and
probability matrices in global memory. This example illustrates how algebraic
reformulation can expose new opportunities for fusion and parallelization.

Prior work has explored both algebraic rewriting~\cite{taso}
and execution optimization through operator scheduling~\cite{tvm}
and fusion~\cite{mirage}. Neptune~\cite{neptune} derives algebraic
repairs that enable the fusion of dependent reductions.
To explore combinations of transformations, equality
saturation~\cite{DBLP:conf/popl/TateSTL09,egg} has been applied
to tensor graph and GPU kernel optimization~\cite{tensat,trinity}.
It represents equivalent expressions in an e-graph, where
successive rewrites can expose further optimization opportunities
before an implementation is selected. Prism~\cite{prism} also
uses e-graph rewriting to verify candidate equivalence during
symbolic program enumeration.

Scaling this exploration to larger programs requires balancing
optimization coverage against search cost. Composable rewrite rules
allow optimizations to combine across operators, creating a large
space of possible implementations. Although e-graphs represent
equivalent expressions compactly, constructing and evaluating
complete candidates becomes increasingly expensive as programs
grow. Limiting the scope of each search reduces this cost, but can
separate operations that would benefit from joint optimization.
TENSAT~\cite{tensat} searches tensor-graph rewrites without exploring
the low-level GPU implementations of individual operators.
Trinity~\cite{trinity} explores loop and memory transformations,
but selects loop structures by kernel count, potentially missing
profitable combinations. The challenge is to preserve these
optimization opportunities while keeping implementation
search tractable.

We present \system, a tensor program superoptimizer that generates
implementation structures through equality saturation. We jointly
design its IR, equality rules, and extraction procedure to make
this search practical.
As shown in Figure~\ref{fig:overview}, our approach explores graph partitions and
generates candidates for each subgraph through equality saturation
and extraction with compaction.  It then composes these candidates into complete
programs and applies code generation and tuning to produce an optimized
implementation. \system is organized around three key designs.

\noindent\textbf{Define the space.}
We introduce a pure expression language that represents high-level
tensor expressions and tiled computations. Both forms coexist in
the same e-graph, allowing subexpressions to be refined while the
rest of the computation remains available for algebraic rewriting.
The IR's types track logical tensor shapes and parallel state,
constraining how tiled computations are constructed and combined.

\noindent\textbf{Search the space.}
We develop composable equality rules for transforming tensor
expressions and their implementations. Parallel refinement
introduces tiling at individual operators. Fusion rules combine
reductions over a shared domain, directly for independent
reductions and with repair functions for dependent reductions.
Propagation rules extend the tiling through connected operations,
allowing these transformations to compose into complex
implementations. \system derives fused
implementations, including FlashAttention-style kernels,
directly from tensor expressions.

\noindent\textbf{Compact the space.}
Equality search produces many candidates whose differences are
unlikely to affect performance, yet constructing and evaluating
them separately is costly. We group such candidates into
equivalence classes and search the resulting quotient space.
Early pruning identifies equivalent partial programs
and selects one representative, avoiding redundant construction
of complete candidates.

We implement \system as a compiler from PyTorch~\cite{pytorch2} programs to
Triton~\cite{triton} kernels, using \textsc{egg}~\cite{egg} for equality
saturation.  Across a range of tensor programs, \system achieves a
$\ProgramSpeedup\times$ geometric mean speedup over the fastest available
baseline, with speedups up to $\ProgramMaxSpeedup\times$.
For attention, \system achieves up to a $\MHADecodeMaxSpeedup\times$
speedup over FlashAttention in decode, with similar performance in prefill.
We further explore optimizations for various
Transformer layers and present five case studies, including QK-normalized
MLA and mHC implementations with $3.16\times$ and $5.84\times$ speedups
over \texttt{torch.compile}, respectively.
Early compaction reduces the number of expanded states by up to a factor
of $2{,}465$ compared with deduplication after completion.

%% file: figures/overview.tex
\begin{figure}[t]
  \centering
  \begingroup
  \definecolor{overviewblue}{HTML}{DCEBFA}
  \definecolor{overviewgreen}{HTML}{DDF1E5}
  \hypersetup{linkcolor=black}
  \newcommand{\overviewsection}[1]{{\normalfont\sffamily
    \fontsize{8}{9.5}\selectfont(\S#1)}}
  \scalebox{0.9}{%
    \begin{tikzpicture}[
     x=1pt,y=-1pt,text=black,font=\sffamily\fontsize{8.5}{10}\selectfont,
     flow/.style={-{Latex[length=2.8pt,width=2.3pt]},draw=black,line width=.55pt},
     dep/.style={-{Latex[length=2.3pt,width=1.9pt]},draw=black,line width=.5pt},
     panel/.style={draw=black,rounded corners=2.5pt,line width=.55pt},
     card/.style={draw=black,fill=white,rounded corners=1.5pt,line width=.45pt},
     title/.style={font=\sffamily\bfseries\fontsize{9}{10.5}\selectfont,inner sep=0pt},
     candidate label/.style={inner sep=0pt,font=\sffamily\fontsize{8.3}{10}\selectfont}
    ]
    \path[use as bounding box] (0,0) rectangle (238,258);
    \node[panel,fill=white,minimum width=158pt,minimum height=17pt,
      title] (input) at (119,8.5) {Input tensor program};
    \draw[flow] (input.south) -- (119,28);
    \node[anchor=west,inner sep=0pt,
      font=\sffamily\fontsize{8}{9.5}\selectfont] at (127,22.5)
      {Partition \overviewsection{\ref{sec:subgraph-composition}}};
    \begin{scope}[shift={(0,13)}]
    
    \filldraw[panel,fill=white] (0,15) rectangle (238,56);
    \foreach \x/\i in {35.5/1,119/2,202.5/3}{
     \node[card,dashed,dash pattern=on 2pt off 1.5pt,
       minimum width=53pt,minimum height=16pt,inner sep=0pt]
       (sub\i) at (\x,35) {Subgraph \i};
    }
    \draw[dep] (sub1.east) -- (sub2.west);
    \draw[dep] (sub2.east) -- (sub3.west);
    \draw[dep] (sub1.south) -- (35.5,49) -- (202.5,49) -- (sub3.south);
    \foreach \x in {35.5,119,202.5}{\draw[flow] (\x,56) -- (\x,66);}
    
    \filldraw[panel,fill=overviewblue] (0,66) rectangle (238,95);
    \node[title] at (119,75) {Equality saturation \overviewsection{\ref{sec:unified-ir}--\ref{sec:equality-rules}}};
    \foreach \x/\i in {35.5/1,119/2,202.5/3}{
     \node[card,minimum width=53pt,minimum height=11pt,inner sep=0pt,
       font=\sffamily\fontsize{8.3}{10}\selectfont]
       at (\x,87) {E-graph $G_{\i}$};
     \draw[flow] (\x,95) -- (\x,105);
    }
    
    \filldraw[panel,fill=overviewgreen] (0,105) rectangle (238,142);
    \node[title] at (119,114) {Extraction with compaction \overviewsection{\ref{sec:extraction-compaction}}};
    \foreach \x/\i in {35.5/1,119/2,202.5/3}{
     \filldraw[card] (\x-22,122) rectangle (\x+27,135);
     \filldraw[card] (\x-24.5,124.5) rectangle (\x+24.5,137.5);
     \filldraw[card] (\x-27,127) rectangle (\x+22,140);
     \node[candidate label] at (\x-2.5,133.5) {Candidate \i};
     \draw[flow] (\x,142) -- (\x,152);
    }
    
    \filldraw[card] (5,152) rectangle (238,191);
    \filldraw[card] (2.5,154.5) rectangle (235.5,193.5);
    \filldraw[card] (0,157) rectangle (233,196);
    \node[title] at (117,166) {Compose subgraph implementations};
    \foreach \x/\i in {35.5/1,119/2,202.5/3}{
     \node[card,minimum width=49pt,minimum height=13pt,inner sep=0pt]
       (impl\i) at (\x-2.5,182) {Candidate \i};
    }
    \draw[dep] (impl1.east) -- (impl2.west);
    \draw[dep] (impl2.east) -- (impl3.west);
    \draw[dep] (impl1.south) -- (33,191.5) -- (200,191.5) -- (impl3.south);
    
    \draw[flow] (119,196) -- (119,205);
    \node[panel,fill=white,minimum width=216pt,minimum height=18pt,
     title] (backend) at (119,214) {Code generation and tuning \overviewsection{\ref{sec:code-generation-and-tuning}}};
    \draw[flow] (backend.south) -- (119,234);
    \node[title,anchor=north] at (119,235) {Optimized tensor program};
    \end{scope}
    \end{tikzpicture}
  }%
  \endgroup
  \caption{Overview of \system.}
  \Description{A compiler overview showing one program partitioned into three
  connected subgraphs. Each column follows one subgraph through equality
  saturation and extraction with compaction. Stacked cards represent candidate
  programs. Composition connects candidates through the original
  subgraph dependencies, including a shared value used by multiple subgraphs,
  to form complete programs. Code generation and tuning produce the
  optimized tensor program. The blocks represent subgraphs and their candidate
  implementations, which may contain multiple kernels.}
  \label{fig:overview}
\end{figure}

%% file: sections/02-equality-saturation.tex
\input{figures/egraph-factoring}

\section{Equality Saturation}
\label{sec:equality-saturation}

An e-graph compactly represents expressions and known equalities between
them~\cite{egg}.  An \emph{e-node} records an operator and references to child
\emph{e-classes}.  Each e-class groups e-nodes representing equivalent
expressions.  For the tensor expression $\operatorname{Exp}(AB+AC)$ in
Figure~\ref{fig:egraph-factoring}, the Exp e-node points to e-class $c$
representing its argument $AB+AC$.

Equality saturation repeatedly applies valid equalities to this
representation~\cite{DBLP:conf/popl/TateSTL09}.  Factoring out $A$ adds
$A(B+C)$ alongside $AB+AC$ in $c$.  The Exp node continues to reference
this class, so the e-graph represents both $\operatorname{Exp}(AB+AC)$ and
$\operatorname{Exp}(A(B+C))$ without duplicating the Exp node.
The e-graph
maintains \emph{congruence}: applying the same operator to equivalent children
produces equivalent expressions.  New expressions can expose further matches,
allowing rules to compose. Rewriting continues until the e-graph is saturated
or a resource limit is reached.

\emph{Extraction} selects a finite expression or program from the e-graph,
even if the e-graph contains cycles.  Starting from the result e-class, it
chooses an e-node and continues resolving choices in its child e-classes until
a complete program is formed.  In the example, choosing the factored MatMul
in $c$ still requires selecting an expression for its $B+C$ operand.
The two forms have the same value but use different numbers of matrix
multiplications, so their execution costs can differ. Extractors may select
programs by minimizing AST size~\cite{egg} or by solving an integer linear
program~\cite{DBLP:journals/pacmpl/GoharshadyLP24}.
\system applies equality saturation to a unified IR of high-level
tensor expressions and tiled computations. Algebraic rewrites and
parallel refinements add equivalent expressions to the same e-graph,
from which extraction selects complete programs for performance
evaluation.

%% file: figures/egraph-factoring.tex
\begin{figure}[t]
  \centering
  \begingroup
  \definecolor{egraphold}{HTML}{DCEBFA}
  \definecolor{egraphnew}{HTML}{DDF1E5}
  \definecolor{egraphnewedge}{HTML}{4F8E78}
  \resizebox{\columnwidth}{!}{%
    \begin{tikzpicture}[
     x=1pt,y=-1pt,text=black,
     font=\sffamily\fontsize{8}{10}\selectfont,
     ec/.style={draw=black,dashed,dash pattern=on 2pt off 1.4pt,
       rounded corners=2pt,line width=.5pt,inner sep=0pt,fill=white},
     en/.style={ellipse,draw=black,line width=.55pt,fill=egraphold,
       minimum width=16pt,minimum height=16pt,inner xsep=3pt,inner ysep=1pt},
     edge/.style={-{Latex[length=2.6pt,width=2.2pt]},line width=.55pt,
       draw=black,preaction={draw=white,line width=1.8pt}},
     newedge/.style={edge,draw=egraphnewedge},
     heading/.style={font=\sffamily\bfseries\fontsize{8.5}{10}\selectfont,
       inner sep=0pt},
     classid/.style={font=\fontsize{8}{10}\selectfont,inner sep=0pt,anchor=east}
    ]
    \path[use as bounding box] (-1,0) rectangle (274,155);
    \node[heading] at (43,5) {Initial e-graph};
    \node[heading] at (203,5) {After factoring};
    
    \node[ec,minimum width=32pt,minimum height=24pt] (lfc) at (43,27) {};
    \node[en] (lf) at (43,27) {Exp};
    \node[ec,minimum width=26pt,minimum height=24pt] (lroot) at (43,58) {};
    \node[en] (ladd) at (43,58) {$+$};
    \node[classid] at (27,49) {$c$};
    \node[ec,minimum width=42pt,minimum height=24pt] (labc) at (21,91) {};
    \node[en,inner xsep=0pt] (lab) at (21,91) {MatMul};
    \node[ec,minimum width=42pt,minimum height=24pt] (lacc) at (69,91) {};
    \node[en,inner xsep=0pt] (lac) at (69,91) {MatMul};
    \foreach \x/\name in {12/A,45/B,78/C}{
     \node[ec,minimum width=24pt,minimum height=24pt] (l\name c) at (\x,127) {};
     \node[en] at (\x,127) {$\name$};
    }
    \draw[edge] (lf.south) -- (lroot.north);
    \draw[edge] (ladd.south west) -- (labc.north);
    \draw[edge] (ladd.south east) -- (lacc.north);
    \draw[edge] (lab.south west) -- ([xshift=-3pt]lAc.north);
    \draw[edge] (lab.south east) -- ([xshift=-3pt]lBc.north);
    \draw[edge] (lac.south west) -- ([xshift=5pt]lAc.north);
    \draw[edge] (lac.south east) -- (lCc.north);
    
    \draw[-{Latex[length=5pt,width=4pt]},line width=1pt]
     (94,59) -- (122,59);
    
    \node[ec,minimum width=32pt,minimum height=24pt] (rfc) at (211,27) {};
    \node[en] (rf) at (211,27) {Exp};
    \node[ec,minimum width=120pt,minimum height=24pt] (rroot) at (211,58) {};
    \node[classid] at (148,49) {$c$};
    \node[en] (radd) at (165,58) {$+$};
    \node[en,fill=egraphnew,draw=egraphnewedge,inner xsep=0pt] (rfactor) at (245,58) {MatMul};
    \node[ec,minimum width=42pt,minimum height=24pt] (rabc) at (141,91) {};
    \node[en,inner xsep=0pt] (rab) at (141,91) {MatMul};
    \node[ec,minimum width=42pt,minimum height=24pt] (racc) at (189,91) {};
    \node[en,inner xsep=0pt] (rac) at (189,91) {MatMul};
    \node[ec,draw=egraphnewedge,minimum width=26pt,minimum height=24pt]
     (rsumc) at (245,91) {};
    \node[en,fill=egraphnew,draw=egraphnewedge] (rsum) at (245,91) {$+$};
    \foreach \x/\name in {143/A,193/B,245/C}{
     \node[ec,minimum width=24pt,minimum height=24pt] (r\name c) at (\x,127) {};
     \node[en] at (\x,127) {$\name$};
    }
    \draw[edge] (rf.south) -- (rroot.north);
    \draw[edge] (radd.south west) -- (rabc.north);
    \draw[edge] (radd.south east) -- (racc.north);
    \draw[edge] (rab.south west) -- ([xshift=-4pt]rAc.north);
    \draw[edge] (rab.south east) -- ([xshift=-5pt]rBc.north);
    \draw[edge] (rac.south west) -- ([xshift=5pt]rAc.north);
    \draw[edge] (rac.south east) -- ([xshift=-4pt]rCc.north);
    \draw[newedge] (rfactor.west)
     -- (219,58) -- (219,74) -- (113,74)
     -- (113,127) -- (rAc.west);
    \draw[newedge] (rfactor.south) -- (rsumc.north);
    \draw[newedge] (rsum.south west) -- ([xshift=5pt]rBc.north);
    \draw[newedge] (rsum.south) -- (rCc.north);
    
    \node[inner sep=0pt] at (43,150) {$\operatorname{Exp}(AB+AC)$};
    \node[inner sep=0pt] at (203,150) {$\operatorname{Exp}\!\left(A(B+C)\right)$};
    \end{tikzpicture}
  }%
  \endgroup
  \caption{Factoring $AB+AC$ within an e-graph. Dashed boxes denote e-classes;
    solid nodes denote e-nodes.}
  \Description{Before factoring, Exp references e-class c, which contains an
  addition of MatMul(A,B) and MatMul(A,C). Factoring adds MatMul(A,B+C) to the
  same e-class and creates an e-class for B+C. The existing nodes remain,
  including the shared input classes A, B, and C. The Exp node still points to
  c, representing both formulations without duplicating the Exp node.
  Blue nodes are existing, and green nodes and edges are additions.
  Every operand edge points to an e-class boundary.}
  \label{fig:egraph-factoring}
\end{figure}
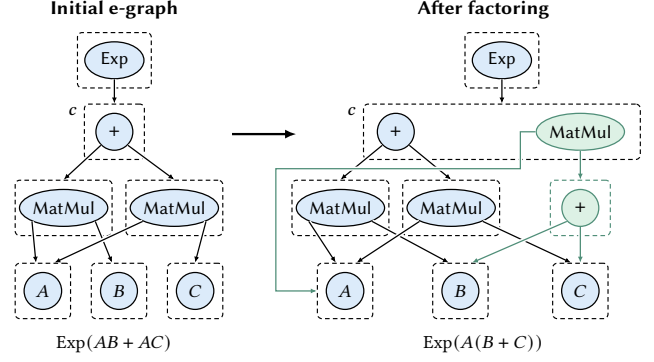

%% file: sections/03-unified-ir.tex
\section{A Unified IR for Tensor Programs}
\label{sec:unified-ir}

We use a single expression language to describe tensor computations
and their parallel implementations.
Figure~\ref{fig:ir} presents the core syntax of our IR.  Alongside tensor
operators, it includes primitives that define parallel operand views, express
reductions, and store intermediate results.  Expressions are pure and produce tensor or tuple values,
with \ir{Tuple} grouping tensor values and \ir{Get} selecting individual fields.
Here, $x$ names a function argument, $c$ is a scalar literal, and $\rho$
denotes a reducer function.

\input{figures/ir-definition}

\subsection{Parallelism}

We first introduce our representation of parallel tensor computations.
Parallel state describes how a tensor expression is represented by values
indexed by a logical parallel axis $p$.  Sharing a parallel axis aligns operand
views and results at corresponding coordinates.
Figure~\ref{fig:distribution-semantics} illustrates parallel decomposition
and the combination of results.

\ir{Partition} defines how a tensor is divided among parallel computations.
\ir{Partition}$_{i,p}$ maps each coordinate of $p$ to a distinct slice
along dimension $i$.  It records $\mathsf{partition}(i,p)$ while preserving the
tensor's logical shape and dtype.
\ir{Combination}$_{i,p}$ joins these slices along $i$ to form the complete
tensor.  It requires the matching state and removes $p$.

\ir{Replicate}$_p$ supplies the same operand to every computation across $p$.
It records $\mathsf{replicate}(p)$, preserving logical and
per-coordinate shapes without storing extra copies.

Partial results express a parallel decomposition of a reduction: each
coordinate computes a contribution to the same output elements.
The state $\mathsf{partial}(p)$ records that these contributions still need to
be combined across $p$.  A parallel \ir{Reduce} combines them into the complete
result and removes $p$ from the state.

By including parallel state in tensor type, we can
distinguish slices, replicas, and partial contributions even when their logical
shapes and dtypes match.  Each e-class has a single value type, so its equivalent
expressions can be used interchangeably by the operations that reference it.

\input{figures/distribution-semantics}

\input{figures/attention-rewrites}
\input{figures/rule-summary}

\subsection{Computation on Tiles}
\label{sec:ir-local-computation}

The parallel views introduced above provide the local operands
for tile computations.
Pointwise operations act independently on each tile and can be fused into
tile computations.  We therefore use the same pointwise operators in
tensor-level and tiled expressions.  Matrix multiplication and reductions
aggregate values along tensor dimensions, so
their tiled forms express how local results are computed and combined.

For matrix multiplication, we use distinct tensor-level and tile-level operators.
\ir{MatMul} represents matrix multiplication as a high-level
tensor operation.
The high-level form allows algebraic
rewrites, such as associativity, to compose with parallel refinement.
\ir{TileMatMul} computes the local product at each parallel coordinate,
producing an output tile depending on the partitioned
dimensions.  Refinement rules relate \ir{MatMul} to equivalent tiled
expressions in the e-graph.  
For example, the following two forms compute a matrix product followed by a
pointwise exponential.  The tiled form uses row tiles along M, indexed by $p$:
\begin{lstlisting}[style=irpaperexample]
// Tensor-level form
a: tensor<MxK,bf16> = input
b: tensor<KxN,bf16> = input
product = matmul(a,b)
output = exp(product)
// Equivalent tiled form
a_tile = partition[M,p](a)
b_tile = replicate[p](b)
product_tile = tile_matmul(a_tile,b_tile)
output_tile = exp(product_tile)
output = combination[M,p](output_tile)
\end{lstlisting}
At each coordinate of $p$, \ir{TileMatMul} multiplies a row tile of \code{a}
by the replicated \code{b}.  The same \ir{Exp} operator applies to the local
product and preserves its partitioned state.  \ir{Combination} then joins the
output tiles.  

\subsection{Reduction State and Intermediate Storage}
\label{sec:ir-reduction-state}

$\ir{Reduce}_{\rho}(e;\alpha)$ applies reducer function $\rho$ to the values of
$e$ along the axis specified by $\alpha$.  The reducer takes the current state
and an input value, and uses IR operations to compute the updated state.
With $\alpha=\mathsf{axis}(i)$, it
reduces tensor dimension $i$, operating within each tile when the input is
partitioned.  With $\alpha=\mathsf{parallel}(p)$, it combines values across
the coordinates of $p$.  The reduction state can be a tensor or a tuple of tensors.

The following example computes a row-wise sum of an $M\times K$ tensor.
In the tiled form, \code{axis=K} computes partial row sums with state
$\mathsf{partial}(p)$, and \code{parallel=p} combines them:
\begin{lstlisting}[style=irpaperexample]
// Tensor-level form
x: tensor<MxK,f32> = input
output = reduce x axis=K reducer=@add
// Equivalent tiled form
x_tile = partition[K,p](x)
partial = reduce x_tile axis=K reducer=@add
output = reduce partial parallel=p reducer=@add
\end{lstlisting}
Partial products from K-partitioned \ir{TileMatMul} are combined by a reduction
over $p$.

Keeping intermediate values within a kernel avoids intermediate memory traffic
but can limit parallelism.  Storing these values in global memory allows
independent thread blocks to produce inputs for a later reduction kernel.
To express this choice in our IR, we introduce \ir{Collection}.
It converts a parallel axis $p$ into a leading tensor dimension of extent $|p|$, preserving one slice
per coordinate and removing $p$ from the parallel state.  For tuple inputs,
this transformation applies to each field.  In the row-sum example, the partial
sums can be stored before the final reduction:
\begin{lstlisting}[style=irpaperexample]
stored = collection[p](partial)
output = reduce stored axis=0 reducer=@add
\end{lstlisting}
The tensor \code{stored} has shape $|p|\times M$, and reducing axis 0 yields
the same row sums.
With \ir{Collection}, we can express fused and multi-kernel
implementations as alternatives in equality search.
Our type system records the different shapes and parallel states
of their intermediate values, while both forms return the same
result type.

%% file: figures/ir-definition.tex
\begin{figure}[!t]
  \centering
  \begingroup
  \begin{tabular}{@{}>{$}r<{$}@{\ }>{$}c<{$}@{\ }>{$}l<{$}@{}}
    \toprule
    \multicolumn{3}{@{}l}{\textsc{Expressions}} \\
    e &::=& \ir{Input}(x) \mid \ir{Constant}(c) \\
      &\mid& \ir{Tuple}(e_1,\ldots,e_n) \mid \ir{Get}_{j}(e) \\
    \addlinespace[2pt]
      &\mid& \ir{Partition}_{i,p}(e)
        \mid \ir{Replicate}_{p}(e) \\
      &\mid& \ir{Combination}_{i,p}(e) \\
    \addlinespace[2pt]
      &\mid& \ir{Exp}(e) \mid e+e \mid e\times e \mid \cdots \\
      &\mid& \ir{MatMul}(e,e) \mid \ir{TileMatMul}(e,e) \\
    \addlinespace[2pt]
      &\mid& \ir{Reduce}_{\rho}(e;\alpha) \\
      &\mid& \ir{Collection}_{p}(e) \\
    \addlinespace
    \multicolumn{3}{@{}l}{\textsc{Types}} \\
    t &::=& \mathsf{Tensor}(\mathsf{dtype},\mathsf{shape},\delta)
      \mid \mathsf{Tuple}(t_1,\ldots,t_n) \\
    \delta &::=& \{d_1,\ldots,d_n\} \\
    d &::=& \mathsf{partition}(i,p) \mid \mathsf{replicate}(p)
      \mid \mathsf{partial}(p) \\
    \addlinespace
    \multicolumn{3}{@{}l}{\textsc{Reduction axes}} \\
    \alpha &::=& \mathsf{axis}(i) \mid \mathsf{parallel}(p) \\
    \bottomrule
  \end{tabular}
  \endgroup

  \caption{Core expressions and parallel-state types.}
  \Description{A compact BNF grammar grouping basic constructs, parallel
  operand views, tensor computations, and reduction and storage constructs
  within one expression category. It also shows tensor and tuple types with
  parallel state, and tensor or parallel reduction axes.}
  \label{fig:ir}
\end{figure}

%% file: figures/distribution-semantics.tex
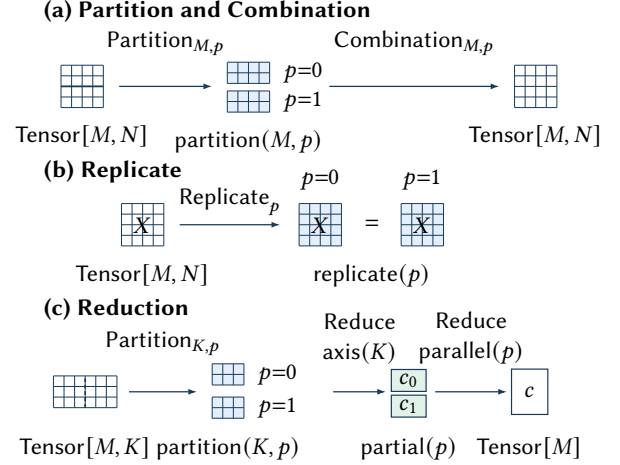
\begin{figure}[t]
  \centering
  \begingroup
  \definecolor{distributionink}{HTML}{193B5A}
  \definecolor{distributionblue}{HTML}{E4EFFB}
  \definecolor{distributiongreen}{HTML}{E1F2E9}
  \newcommand{\efgrid}[6]{%
    \begin{scope}[shift={(#1,#2)}]
      \pgfmathsetmacro{\gw}{#3*0.14}
      \pgfmathsetmacro{\gh}{#4*0.14}
      \fill[#5] (0,0) rectangle (\gw,\gh);
      \foreach \gx in {0,...,#3}
        \draw[#6,line width=0.28pt] ({\gx*0.14},0)--({\gx*0.14},\gh);
      \foreach \gy in {0,...,#4}
        \draw[#6,line width=0.28pt] (0,{\gy*0.14})--(\gw,{\gy*0.14});
    \end{scope}%
  }
  \begin{tikzpicture}[
      x=1cm,y=1cm,text=black,
      flow/.style={-{Latex[length=1.15mm,width=0.8mm]},
        draw=distributionink,line width=0.48pt},
      opname/.style={font=\small\sffamily,text=black,
        fill=white,inner xsep=1.5pt,inner ysep=0.5pt,
        text height=7pt,text depth=3pt},
      rowname/.style={font=\small\sffamily\bfseries,text=black,
        anchor=west},
      typelabel/.style={font=\small\sffamily,text=black,
        anchor=north,align=center},
      plabel/.style={font=\small\sffamily,text=black,anchor=west}]

    \node[rowname] at (0,4.55) {(a) Partition and Combination};
    \efgrid{0.35}{3.28}{4}{4}{white}{distributionink}
    \draw[distributionink,line width=0.65pt]
      (0.35,3.56)--(0.91,3.56);
    \node[typelabel] at (0.63,3.17) {$\mathsf{Tensor}[M,N]$};

    \draw[flow] (1.12,3.56)--(2.35,3.56)
      node[midway,opname,above=11pt] {$\ir{Partition}_{M,p}$};
    \efgrid{2.55}{3.60}{4}{2}{distributionblue}{distributionink}
    \efgrid{2.55}{3.22}{4}{2}{distributionblue}{distributionink}
    \node[plabel] at (3.18,3.74) {$p{=}0$};
    \node[plabel] at (3.18,3.36) {$p{=}1$};
    \node[typelabel,text=black] at (2.83,3.10)
      {$\mathsf{partition}(M,p)$};

    \draw[flow] (3.90,3.56)--(6.15,3.56)
      node[midway,opname,above=11pt] {$\ir{Combination}_{M,p}$};
    \efgrid{6.35}{3.28}{4}{4}{white}{distributionink}
    \node[typelabel] at (6.63,3.17) {$\mathsf{Tensor}[M,N]$};

    \begin{scope}[shift={(0,-0.25)}]
    \node[rowname] at (0,2.68) {(b) Replicate};
    \efgrid{1.15}{1.70}{4}{4}{white}{distributionink}
    \node[font=\small\sffamily,text=black] at (1.43,1.98) {$X$};
    \node[typelabel] at (1.43,1.59) {$\mathsf{Tensor}[M,N]$};

    \draw[flow] (1.92,1.98)--(3.25,1.98)
      node[midway,opname,above=4pt] {$\ir{Replicate}_{p}$};
    \efgrid{3.50}{1.70}{4}{4}{distributionblue}{distributionink}
    \node[font=\small\sffamily,text=black] at (3.78,1.98) {$X$};
    \node[plabel,anchor=south] at (3.78,2.32) {$p{=}0$};
    \node[font=\small\sffamily,text=black] at (4.46,1.98) {$=$};
    \efgrid{4.85}{1.70}{4}{4}{distributionblue}{distributionink}
    \node[font=\small\sffamily,text=black] at (5.13,1.98) {$X$};
    \node[plabel,anchor=south] at (5.13,2.32) {$p{=}1$};
    \node[typelabel,text=black] at (4.46,1.59)
      {$\mathsf{replicate}(p)$};
    \end{scope}

    \begin{scope}[shift={(0,-0.78)}]
    \node[rowname] at (0,1.40) {(c) Reduction};
    \efgrid{0.25}{0.09}{6}{3}{white}{distributionink}
    \draw[distributionink,line width=0.65pt,dashed,dash pattern=on 1.4pt off 1pt]
      (0.67,0.09)--(0.67,0.51);
    \node[typelabel] at (0.67,-0.17) {$\mathsf{Tensor}[M,K]$};

    \draw[flow] (1.25,0.30)--(2.15,0.30)
      node[midway,opname,above=13pt] {$\ir{Partition}_{K,p}$};
    \efgrid{2.35}{0.39}{3}{2}{distributionblue}{distributionink}
    \efgrid{2.35}{-0.05}{3}{2}{distributionblue}{distributionink}
    \node[plabel] at (2.84,0.53) {$p{=}0$};
    \node[plabel] at (2.84,0.09) {$p{=}1$};
    \node[typelabel,text=black] at (2.56,-0.17)
      {$\mathsf{partition}(K,p)$};

    \draw[flow] (3.95,0.30)--(4.65,0.30)
      node[midway,opname,above=10pt,align=center,text height={},text depth={}]
        {\ir{Reduce}\\$\mathsf{axis}(K)$};
    \node[draw=distributionink,fill=distributiongreen,line width=0.45pt,
      minimum width=4.6mm,minimum height=2.9mm,inner sep=0pt,
      font=\small\sffamily] at (4.95,0.45) {$c_0$};
    \node[draw=distributionink,fill=distributiongreen,line width=0.45pt,
      minimum width=4.6mm,minimum height=2.9mm,inner sep=0pt,
      font=\small\sffamily] at (4.95,0.11) {$c_1$};
    \node[typelabel,text=black] at (4.95,-0.17)
      {$\mathsf{partial}(p)$};

    \draw[flow] (5.30,0.30)--(6.25,0.30)
      node[midway,opname,above=10pt,align=center,text height={},text depth={}]
        {\ir{Reduce}\\$\mathsf{parallel}(p)$};
    \node[draw=distributionink,fill=white,line width=0.45pt,
      minimum width=4.8mm,minimum height=6.0mm,inner sep=0pt,
      font=\small\sffamily] at (6.55,0.30) {$c$};
    \node[typelabel] at (6.55,-0.17) {$\mathsf{Tensor}[M]$};
    \end{scope}
  \end{tikzpicture}
  \endgroup

  \caption{Parallelism transitions: (a) partition and combination,
  (b) replication, and (c) reduction.}
  \Description{Three compact tensor-grid diagrams.  The first splits a matrix
  along M and combines the two tiles.  The second creates two identical matrix
  replicas.  The third partitions tensor dimension K along parallel axis p,
  reduces K within each slice to obtain a partial value, and reduces p to
  combine the partials into one complete tensor.}
  \label{fig:distribution-semantics}
\end{figure}

%% file: figures/attention-rewrites.tex
\begin{figure*}[t]
  \centering
  \begingroup
  \newcommand{\figop}[1]{\textcolor{paperblue!78!black}{\bfseries #1}}
  \newcommand{\figpar}[1]{\textcolor{paperteal!92!black}{\bfseries #1}}
  \newcommand{\figattr}[1]{\textcolor{paperteal!92!black}{#1}}
  \newcommand{\figtype}[1]{\textcolor{paperpurple}{#1}}
  \makebox[\textwidth][c]{%
  \begin{tikzpicture}[
      x=1cm,y=1cm,
      code line/.style={anchor=west,inner xsep=0pt,inner ysep=.22pt,
        font=\small\ttfamily,text=black,
        text height=1.55ex,text depth=.38ex},
      code matrix/.style={matrix of nodes,anchor=north west,
        nodes in empty cells,inner sep=0pt,row sep=.22pt,column sep=0pt,
        column 1/.style={nodes={code line}}},
      compact gap/.style={inner sep=0pt,text height=.50ex,text depth=0pt},
      region gap/.style={inner sep=0pt,text height=1.10ex,text depth=0pt},
      form title/.style={anchor=south,inner sep=0pt,
        font=\small\sffamily\bfseries,text=paperink},
      parallel title/.style={anchor=south west,inner sep=0pt,
        font=\small\sffamily\bfseries,text=paperteal!92!black},
      repair title/.style={anchor=south west,inner sep=0pt,
        font=\small\sffamily\bfseries,text=paperpurple!82!black},
      repair bar/.style={draw=paperpurple!78!black,line width=1.05pt},
      repair arrow/.style={-{Latex[length=1.12mm,width=.78mm]},
        draw=paperpurple!76!black,line width=.48pt},
      parallel arrow/.style={-{Latex[length=1.12mm,width=.78mm]},
        draw=paperteal!88!black,line width=.48pt},
      green region/.style={draw=paperteal!82!black,line width=.46pt,
        rounded corners=.55mm,inner xsep=1.05mm,inner ysep=.25mm},
      purple region/.style={draw=paperpurple!74!black,line width=.46pt,
        rounded corners=.55mm,inner xsep=1.05mm,inner ysep=.25mm},
      rule label/.style={font=\footnotesize\sffamily,align=center,
        fill=white,inner xsep=1.2pt,inner ysep=.30pt}]

    \path (-.105,0) node[inner sep=0pt] {};
    \matrix (src) [code matrix] at (.25,0) {
      |[name=sq]| q: \figtype{tensor<QxD,bf16>} = \figop{input} \\
      |[name=sk]| k: \figtype{tensor<KxD,bf16>} = \figop{input} \\
      |[name=sv]| v: \figtype{tensor<KxD,bf16>} = \figop{input} \\
      |[name=sscale]| scale: \figtype{tensor<1,f32>} = \figop{constant} \\
      |[name=sgap0,compact gap]| {} \\
      |[name=skt]| k\_t = \figop{transpose}[1,0](k) \\
      |[name=sscore]| score = \figop{matmul}(q,k\_t) \figattr{acc=f32} \\
      |[name=sscaled]| scaled = \figop{multiply}(score,scale) \\
      |[name=sgap1,region gap]| {} \\
      |[name=smax]| maximum = \figop{reduce} scaled
        \figattr{axis=K reducer=@max} \\
      |[name=sweights]| weights = \figop{exp}(\figop{subtract}(scaled,maximum)) \\
      |[name=snorm]| normalizer = \figop{reduce} weights
        \figattr{axis=K reducer=@add} \\
      |[name=sgap2,region gap]| {} \\
      |[name=sprob]| prob = \figop{divide}(weights,normalizer) \\
      |[name=sout]| output = \figop{matmul}(prob,v) \figattr{acc=f32} \\
    };
    \matrix (dst) [code matrix] at (10.55,0) {
      |[name=tq]| q\_tile = \figpar{replicate}[p](\figpar{partition}[Q,m](q)) \\
      |[name=tk]| k\_tile = \figpar{partition}[K,p](\figpar{replicate}[m](k)) \\
      |[name=tv]| v\_tile = \figpar{partition}[K,p](\figpar{replicate}[m](v)) \\
      |[name=tscale]| scale\_tile = \figpar{replicate}[p](\figpar{replicate}[m](scale)) \\
      |[name=tgapfront,compact gap]| {} \\
      |[name=tkt]| k\_t\_tile = \figop{transpose}[1,0](k\_tile) \\
      |[name=tscore]| score = \figop{tile\_matmul}(q\_tile,k\_t\_tile) \figattr{acc=f32} \\
      |[name=tscaled]| scaled = \figop{multiply}(score,scale\_tile) \\
      |[name=tblockmax]| block\_max = \figop{reduce} scaled
        \figattr{axis=K reducer=@max} \\
      |[name=titem]| item = \figop{tuple}(block\_max,scaled,v\_tile) \\
      |[name=tgap0,compact gap]| {} \\
      |[name=rsig]| \figop{rolling}(state=<max,sum,value>, \\
      |[name=ritem]| \phantom{rolling(}item=<block\_max,scaled,v\_tile>) \{ \\
      |[name=rmax]| \quad next\_max = \figop{maximum}(max,block\_max) \\
      |[name=rrescale]| \quad rescale = \figop{exp}(max-next\_max) \\
      |[name=rweights]| \quad weights = \figop{exp}(scaled-next\_max) \\
      |[name=rsum]| \quad tile\_sum = \figop{reduce} weights
        \figattr{axis=K reducer=@add} \\
      |[name=rnextsum]| \quad next\_sum = rescale*sum + tile\_sum \\
      |[name=tgapbranch,compact gap]| {} \\
      |[name=rvalue]| \quad tile\_value = \figop{tile\_matmul}(weights,v\_tile) \\
      |[name=rnextvalue]| \quad next\_value = rescale*value + tile\_value \\
      |[name=rreturn]| \quad \figop{return}
        \figop{tuple}(next\_max,next\_sum,next\_value) \\
      |[name=rclose]| \} \\
      |[name=tgap1,compact gap]| {} \\
      |[name=tsummary]| summary = \figpar{reduce} item
        \figattr{parallel=p reducer=@rolling} \\
      |[name=tnorm]| normalizer = \figop{get}(summary,sum) \\
      |[name=tnum]| numerator = \figop{get}(summary,value) \\
    };
    \node[form title,anchor=south west] at ($(dst.north west)+(0,.12)$)
      {M2: tiled rolling attention};
    \node[purple region,
      fit=(rsig)(ritem)(rmax)(rrescale)(rweights)(rsum)(rvalue)
        (rnextsum)(rnextvalue)(rreturn)(rclose)(tsummary)(tnorm)(tnum)]
      (rollbox) {};
    \coordinate (m1origin) at ($(src.south west)+(0,-.78)$);
    \matrix (m1) [code matrix] at (m1origin) {
      |[name=m1scaled]| scaled\_tile = \figpar{partition}[K,p](scaled) \\
      |[name=m1v]| v\_tile = \figpar{partition}[K,p](v) \\
      |[name=m1block]| block\_max = \figop{reduce} scaled\_tile
        \figattr{axis=K reducer=@max} \\
      |[name=m1max]| maximum = \figpar{reduce} block\_max
        \figattr{parallel=p reducer=@max} \\
      |[name=m1maxtile]| maximum\_tile = \figpar{replicate}[p](maximum) \\
      |[name=m1weights]| weights = \figop{exp}(scaled\_tile-maximum\_tile) \\
      |[name=m1sum]| tile\_sum = \figop{reduce} weights
        \figattr{axis=K reducer=@add} \\
      |[name=m1value]| tile\_value = \figop{tile\_matmul}(weights,v\_tile) \\
      |[name=m1norm]| normalizer = \figpar{reduce} tile\_sum
        \figattr{parallel=p reducer=@add} \\
      |[name=m1num]| numerator = \figpar{reduce} tile\_value
        \figattr{parallel=p reducer=@add} \\
    };
    \node[form title,anchor=south west]
      at ($(m1.north west)+(0,.12)$)
      {M1: dependent reductions};
    \node[fit=(m1)(rollbox),inner sep=0pt,outer sep=0pt]
      (reduction-cores) {};
    \coordinate (tail-line-y) at ($(reduction-cores.south)+(0,-.29)$);
    \node[code line] (m1out) at (m1.west |- tail-line-y)
      {output = numerator / normalizer};
    \node[code line,anchor=south west] (tout)
      at ($(dst.west |- tail-line-y)+(0,.11pt)$)
      {output = numerator / normalizer};
    \node[code line,anchor=north west] (tresult)
      at ($(dst.west |- tail-line-y)+(0,-.11pt)$)
      {result = \figpar{combination}[Q,m](output)};
    \coordinate (m1repairbar) at ($(m1.east)+(.12,0)$);
    \draw[repair bar] (m1repairbar |- m1.north) --
      (m1repairbar |- m1num.south);

    \coordinate (left-column-content-right)
      at ($(m1repairbar)+(-1.05mm,0)$);
    \coordinate (front-width) at (left-column-content-right |- sscaled.east);
    \node[fit=(sq)(sk)(sv)(sscale)(skt)(sscore)(sscaled)(front-width),
      inner xsep=1.05mm,inner ysep=.25mm]
      (sfrontcore) {};
    \coordinate (source-width) at (left-column-content-right |- sout.east);
    \node[purple region,
      fit=(smax)(sweights)(snorm)(sprob)(sout)(source-width)]
      (sourcecore) {};
    \node[form title,anchor=south west]
      at (src.north west |- {$(sfrontcore.north)+(0,.12)$})
      {M0: vanilla attention};

    \coordinate (expose-start) at ($(sourcecore.south)+(0,-.06)$);
    \coordinate (expose-end) at ($(sourcecore.south |- m1.north)+(0,.20)$);
    \draw[parallel arrow] (expose-start) -- (expose-end);
    \node[rule label,text=paperteal!90!black,anchor=west]
      at ($(expose-start)!0.50!(expose-end)+(.16,0)$)
      {factor motion + K refinement};

    \coordinate (front-source) at ($(skt.east)+(.10,-.10)$);
    \coordinate (front-target) at ($(tkt.west)+(-.14,-.10)$);
    \draw[parallel arrow] (front-source) -- (front-target);
    \node[rule label,text=paperteal!90!black,anchor=south]
      at ($(front-source)!0.50!(front-target)+(0,.05)$)
      {Q/K refinement + propagation};

    \coordinate (joint-source) at (m1repairbar |- rnextvalue.west);
    \coordinate (joint-target) at (rollbox.west |- rnextvalue.west);
    \draw[repair arrow] (joint-source) -- (joint-target);
    \node[rule label,text=paperpurple!82!black,anchor=south]
      at ($(joint-source)!0.50!(joint-target)+(0,.16)$)
      {repair fusion\\$h_{\mathrm{sum}},\ h_{\mathrm{value}}$};

    \coordinate (tail-source-x) at ($(m1out.east)+(.10,0)$);
    \coordinate (tail-target-x) at ($(tout.west)+(-.14,0)$);
    \coordinate (tail-source) at (tail-source-x |- tail-line-y);
    \coordinate (tail-target) at (tail-target-x |- tail-line-y);
    \draw[parallel arrow] (tail-source) -- (tail-target);
    \node[rule label,text=paperteal!90!black,anchor=north]
      at ($(tail-source)!0.50!(tail-target)+(0,-.02)$)
      {propagation + combination};
  \end{tikzpicture}%
  }
  \endgroup

  \caption{Composing repair, refinement, and propagation rules from
  vanilla (M0) to tiled rolling attention (M2). Parallel axes $m$ and $p$
  index tiles along the query dimension Q and key dimension K, respectively.}
  \Description{A two-column derivation shows vanilla attention and the
  K-tiled dependent reductions below it on the left, and the complete tiled
  rolling implementation on the right.  Factor motion, K refinement, and
  propagation expose the intermediate form M1.  The repairs for the sum and
  value are combined in the three-field rolling reducer in M2.
  Green marks refinement, propagation, and combination.
  Purple marks dependent-reduction repair.  Precision conversions are omitted.}
  \label{fig:attention-rewrites}
\end{figure*}

%% file: figures/rule-summary.tex
\begin{table*}[t]
  \caption{Core equality rules for tensor programs.}
  \label{tab:rules}
  \normalsize
\begin{tabularx}{\linewidth}{@{}L{4.2cm}Y@{}}
    \toprule
    \emph{Rule} & \emph{Equality} \\
    \midrule
    \multicolumn{2}{@{}l}{\textbf{Algebraic equalities}} \\
    \addlinespace
    Assoc. and comm.
      & $(a\circ b)\circ c \equiv a\circ(b\circ c)
          \;(\circ\in\{+,\times\})\qquad
          a\star b \equiv b\star a
          \;(\star\in\{+,\times,\max,\min\})$
      \\
    Reassociate \ir{MatMul}
      & $\ir{MatMul}(\ir{MatMul}(A,B),C)
          \equiv \ir{MatMul}(A,\ir{MatMul}(B,C))$
      \\
    Factor motion$^\ast$
      & $\ir{MatMul}(X\circ a,W)\equiv\ir{MatMul}(X,W)\circ a
          \qquad
          \ir{MatMul}(X,W\circ a)\equiv\ir{MatMul}(X,W)\circ a$
      \\
\end{tabularx}
\par
\begin{tabularx}{\linewidth}{@{}L{36pt}L{134pt}L{134pt}Y@{}}
\midrule
\multicolumn{4}{@{}p{\linewidth}@{}}{\textbf{Parallel refinement}\hfill\S\ref{sec:rules-parallel-implementations}}\\
\addlinespace
\multicolumn{4}{@{}l}{$\ir{MatMul}(A,B)\equiv\mathrm{Result};\quad
  A\!:\!\cdots\times M\times K,\ B\!:\!\cdots\times K\times N;\quad
  T=\ir{TileMatMul}(A',B')$.}\\
\emph{Axes} & \emph{Operand $A'$} & \emph{Operand $B'$} & \emph{Result}\\
\cmidrule(lr){1-1}\cmidrule(lr){2-2}\cmidrule(lr){3-3}\cmidrule(l){4-4}
None & $A$ & $B$ & $T$\\
M & $\ir{Partition}_{M,p_M}(A)$
  & $\ir{Replicate}_{p_M}(B)$
  & $\ir{Combination}_{M,p_M}(T)$\\
N & $\ir{Replicate}_{p_N}(A)$
  & $\ir{Partition}_{N,p_N}(B)$
  & $\ir{Combination}_{N,p_N}(T)$\\
MN & $\ir{Replicate}_{p_N}(\ir{Partition}_{M,p_M}(A))$
  & $\ir{Partition}_{N,p_N}(\ir{Replicate}_{p_M}(B))$
  & $\ir{Combination}_{M,p_M}(\ir{Combination}_{N,p_N}(T))$\\
MK & $\ir{Partition}_{K,p_K}(\ir{Partition}_{M,p_M}(A))$
  & $\ir{Partition}_{K,p_K}(\ir{Replicate}_{p_M}(B))$
  & $\ir{Combination}_{M,p_M}(\ir{Reduce}_{+}(T;\mathsf{parallel}(p_K)))$\\
NK & $\ir{Partition}_{K,p_K}(\ir{Replicate}_{p_N}(A))$
  & $\ir{Partition}_{K,p_K}(\ir{Partition}_{N,p_N}(B))$
  & $\ir{Combination}_{N,p_N}(\ir{Reduce}_{+}(T;\mathsf{parallel}(p_K)))$\\
MNK
  & $\begin{aligned}[t]
       &\ir{Partition}_{K,p_K}(\ir{Replicate}_{p_N}(\\
       &\quad\ir{Partition}_{M,p_M}(A)))
     \end{aligned}$
  & $\begin{aligned}[t]
       &\ir{Partition}_{K,p_K}(\ir{Partition}_{N,p_N}(\\
       &\quad\ir{Replicate}_{p_M}(B)))
     \end{aligned}$
  & $\begin{aligned}[t]
       &\ir{Combination}_{M,p_M}(\ir{Combination}_{N,p_N}(\\
       &\quad\ir{Reduce}_{+}(T;\mathsf{parallel}(p_K))))
     \end{aligned}$\\
Batch $b$
  & $\ir{Partition}_{b,p_b}(A)$
  & $\ir{Partition}_{b,p_b}(B)$
  & $\ir{Combination}_{b,p_b}(\ir{MatMul}(A',B'))$\\
\end{tabularx}
\par\smallskip
\begin{tabularx}{\linewidth}{@{}L{36pt}Y@{}}
    \multicolumn{2}{@{}l}{$\ir{Reduce}_{\rho}(X;\mathsf{axis}(K))\equiv\mathrm{Result};
      \quad X\!:\!M\times K$.}\\
    \addlinespace
    M & $\ir{Combination}_{M,p}\!\left(
            \ir{Reduce}_{\rho}(\ir{Partition}_{M,p}(X);
            \mathsf{axis}(K))\right)$\\
    K & $\ir{Reduce}_{\rho}\!\left(
            \ir{Reduce}_{\rho}(\ir{Partition}_{K,p}(X);
            \mathsf{axis}(K));\mathsf{parallel}(p)\right)$\\
\end{tabularx}
\par
\begin{tabularx}{\linewidth}{@{}L{4.2cm}Y@{}}
    \midrule
    \multicolumn{2}{@{}p{\linewidth}@{}}{\textbf{Reduction fusion}\hfill\S\ref{sec:rules-reduction-fusion}}\\
    \addlinespace
    Product fusion
      & $\langle\ir{Reduce}_{\rho_A}(X;\alpha),
          \ir{Reduce}_{\rho_B}(Y;\alpha)\rangle
          \equiv\ir{Reduce}_{\rho_A\times\rho_B}
          (\langle X,Y\rangle;\alpha)$
      \\
    Repair fusion
      & $s=\ir{Reduce}_{\rho_P}(X;\alpha):\quad
          \langle s,\ir{Reduce}_{\rho_C}(g(Y,s);\alpha)\rangle
          \equiv\ir{Reduce}_{\rho_h}(\langle X,Y\rangle;\alpha)$
      \\
    \midrule
    \multicolumn{2}{@{}p{\linewidth}@{}}{\textbf{Propagating parallelism}\hfill\S\ref{sec:rules-propagation}}\\
    \addlinespace
    Pointwise
      & $\begin{array}[t]{@{}l@{}}
          f(\ir{Combination}_{i,p}(x))\equiv
            \ir{Combination}_{i,p}(f(x))\\
          \ir{Partition}_{i,p}(f(x))\equiv
            f(\ir{Partition}_{i,p}(x))\qquad
          \ir{Replicate}_{p}(f(x))\equiv
            f(\ir{Replicate}_{p}(x))
        \end{array}$
      \\
    Replication across reduction
      & $\ir{Replicate}_{q}\bigl(
          \ir{Reduce}_{\rho}(X;\mathsf{parallel}(P))\bigr)
          \equiv
          \ir{Reduce}_{\rho}\bigl(
          \ir{Replicate}_{q}(X);\mathsf{parallel}(P)\bigr),\quad q\notin P$
      \\
    \midrule
    \textbf{Intermediate storage}
      & $\ir{Reduce}_{\rho}(X;\mathsf{parallel}(p))\equiv
          \ir{Reduce}_{\rho}\!\left(
          \ir{Collection}_{p}(X);\mathsf{axis}(0)\right)$\hfill\S\ref{sec:rules-materialization}
      \\
    \bottomrule
\end{tabularx}
\par\smallskip
{\raggedright $^\ast\,\circ\in\{\times,/\}$. The factor $a$ is invariant along K and broadcasts
compatibly. Division requires $a\ne0$.\par}
\end{table*}

%% file: sections/04-equality-rules.tex
\section{Equality Rules for Tensor Programs}
\label{sec:equality-rules}

We now introduce the equality rules that \system uses to explore equivalent
formulations and implementations of tensor programs.
Our algebraic rewrites assume real arithmetic. Floating-point implementations
may differ due to reassociation and internal precision choices.
Figure~\ref{fig:attention-rewrites} illustrates the derivation of tiled rolling
attention (M2) from its tensor-level formulation (M0).  An algebraic rewrite
moves division by the normalizer after the value matrix multiplication,
exposing a weighted-sum numerator.
Refinement and propagation express the computation on tiles,
but dependencies between the reductions prevent direct fusion.
Repair fusion combines these dependent reductions into one,
rescaling the accumulated normalizer and weighted sum
as the running maximum changes.
Table~\ref{tab:rules} summarizes the core equality rules.  

\subsection{Constructing Parallel Implementations}
\label{sec:rules-parallel-implementations}

Parallel refinement expresses tensor-level operators as computations on tiles
together with the operations that combine their results.  For \ir{MatMul},
Table~\ref{tab:rules} specifies the operand views $A',B'$ supplied to
\ir{TileMatMul} for each choice of partitioned dimensions.  Partitioning M or N
produces disjoint output tiles joined by \ir{Combination}. Partitioning K
produces partial products combined by \ir{Reduce}.  Multiple dimensions can be
partitioned together.

Batch refinement partitions each operand along the batch dimension
corresponding to $b$, using \ir{Replicate} when that operand is broadcast along
$b$.  The refined expression contains a tensor-level \ir{MatMul} inside
\ir{Combination}, allowing batch partitioning to compose with the M/N/K
refinements.

Reductions can likewise be refined along an output dimension or a reduction
dimension.  For an $M\times K$ tensor reduced along K, partitioning M produces
complete row results joined by \ir{Combination}.  Partitioning K instead
produces partial row results combined by a parallel \ir{Reduce}.  This second
rule requires an associative reducer that can combine its own partial results,
with both local and final reductions initialized to the identity.

In Figure~\ref{fig:attention-rewrites}, refinement creates tiles along the query
dimension Q and key dimension K, indexed by $m$ and $p$, respectively.  M1 shows
three reductions over $p$, combining tile maxima, partial normalizers, and
partial weighted sums.  The normalizer and weighted sum still depend on the
maximum over all key tiles.

\subsection{Composing Reduction State}
\label{sec:rules-reduction-fusion}

Reduction fusion combines reductions over the same domain into one
reduction with a tuple of their states.  Two reductions are independent if
neither uses the other's result.  Their state updates can then be combined
without modification.  When one reduction's contributions
depend on another reduction's result, fusion requires a repair that adjusts the
accumulated consumer state as the producer state changes.

\emph{Independent state.}
Product fusion pairs the initial states of the two reductions.  For each input
pair $(x,y)$, the product reducer $\rho_A\times\rho_B$ updates the first component
using $\rho_A$ and $x$, and the second using $\rho_B$ and $y$.  The final tuple
contains the results of both original reductions.

\emph{Dependent state.}
Neptune's repair method~\cite{neptune} lets a consumer accumulate contributions
using the producer's current state.  When the producer state changes from $s$ to $s'$, the repair
function $h(c,s,s')$ adjusts the accumulated consumer state $c$
to reflect the new producer state.  For each input $y$ in $Y$, $g(y,s)$ computes its contribution
to the consumer reduction under producer state $s$. $g(Y,s)$
denotes all such contributions. The repair
must update each contribution and distribute over the
consumer reducer $\rho_C$:
\begin{align*}
  h(g(y,s),s,s') &= g(y,s'),\\
  h(\rho_C(a,b),s,s') &=
    \rho_C\!\left(h(a,s,s'),h(b,s,s')\right).
\end{align*}
The fused reducer $\rho_h$ maps state $\langle s,c\rangle$
and input $\langle x,y\rangle$ to $\langle s',c'\rangle$:
\begin{equation*}
  s'=\rho_P(s,x),\qquad
  c'=\rho_C\!\left(h(c,s,s'),\,g(y,s')\right).
\end{equation*}
We represent the fused computation as one \ir{Reduce} over a tuple of producer
and consumer states.  Its reducer shares the producer update and applies each
consumer's repair, while \ir{Get} projects the results.  Because \ir{Reduce}
ranges over either tensor dimensions or parallel axes, the same representation
handles tensor-level and tiled expressions.

In Figure~\ref{fig:attention-rewrites}, the producer state is the running
maximum, which is invariant along K.  The repair
$h(c,s,s')=c\exp(s-s')$ therefore applies to both the accumulated normalizer
and weighted sum.  Combined with refinement and propagation, this produces the
online softmax update~\cite{DBLP:journals/corr/abs-1805-02867} in M2.

\subsection{Propagating Parallelism}
\label{sec:rules-propagation}

To combine the preceding transformations into a tiled implementation, their
producer and consumer operations must use compatible tile views.
Propagation rules extend a chosen decomposition through these operations,
avoiding a specialized rewrite for every operator sequence.

For pointwise operations, the equalities in Table~\ref{tab:rules} move
\ir{Partition} and \ir{Replicate} to the operation's inputs, or move
\ir{Combination} after the operation.  Multiple-input operations pair tiles
at the same parallel coordinate, and broadcast operands use \ir{Replicate}.
Replication can also move across a reduction when the replicated
axis is not reduced.

In Figure~\ref{fig:attention-rewrites}, these equalities propagate the query
and key decompositions through scaling, exponentiation, and transpose.
Together with matrix-multiplication refinement, they express score tiles
directly in terms of query and key tiles, providing inputs to the rolling
reduction in M2.  The final division likewise computes each normalized output
tile before \ir{Combination} joins the results.

\subsection{Storing Reduction Inputs}
\label{sec:rules-materialization}

The storage choice introduced in \S\ref{sec:ir-reduction-state}
can increase parallelism when a fused kernel has few output
tiles, as in decoding with small batches. It lets more thread
blocks produce the reduction inputs independently, at the cost
of intermediate memory traffic and a separate reduction kernel.
The rule in Table~\ref{tab:rules} inserts \ir{Collection} before
\ir{Reduce}, replacing a parallel-axis reduction with a
tensor-axis reduction while preserving the input sequence
and reducer. In M2 of Figure~\ref{fig:attention-rewrites},
these inputs are tuples containing scores, local maxima, and
value tiles. A subsequent kernel consumes the stored tuples
using the same rolling reducer.

%% file: sections/05-scalable-program-extraction.tex
\section{Scalable Program Extraction}
\label{sec:scalable-program-extraction}

The equality rules in the previous section define a space of functionally
equivalent implementations.  \system extracts complete programs from this
space for performance evaluation, fixing their computation and kernel
boundaries while leaving scheduling parameters symbolic.
During extraction, we skip tensor-level \ir{MatMul} nodes
and select their tiled implementations for code generation.
Each program selects
one e-node for every reachable e-class and forms a finite acyclic graph.  We first describe how
compaction avoids redundant extraction, then show how subgraph
implementations are combined to handle larger graphs.

\input{figures/extraction-quotient}

\subsection{Compacting the Search Space}
\label{sec:extraction-compaction}

Extracting and evaluating candidate programs separately can be costly
even when they are expected to perform similarly.
We express this redundancy through an equivalence relation $\approx$ on
candidate programs and explore the resulting quotient space, evaluating one
representative per class.
We group candidates that differ in commutative pointwise operand order
or in how they express the same parallel views.
Candidates with different matrix-multiplication structures, reduction
organizations, or intermediate storage remain distinct.
Figure~\ref{fig:extraction-quotient} groups $S_2=AB+AC$ with
$S_3=AC+AB$, while keeping $S_1=A(B+C)$ distinct because it
uses one matrix multiplication instead of two.

A straightforward approach is to enumerate complete programs and then group
them into equivalence classes. Although this reduces the number of performance
evaluations, extraction must still construct every candidate. Independent
implementation choices can generate exponentially many candidates, making
extraction prohibitively expensive.

To prune redundant branches early, we compare partial programs
containing selected operations and named holes $\square_c$ for
unresolved e-classes $c$. We write $S\approx_p T$ when the two
partial programs allow the same valid assignments to their holes
and each assignment produces complete programs equivalent under
$\approx$. For this comparison, we canonicalize pointwise
expressions and parallel views. For example, in
Figure~\ref{fig:extraction-quotient}, completing
$\square_P+\square_Q$ and $\square_Q+\square_P$ with the same
choices for $P$ and $Q$ produces programs that differ only in
the addition's operand order. One branch can therefore be
pruned before either choice is resolved.

\begin{algorithm}
  \caption{Extraction with early compaction}
  \label{alg:tensor-program-extraction}
  \DontPrintSemicolon
  \KwIn{Typed e-graph $G$, result e-classes $R$, relation $\approx$}
  \KwOut{Representative parameterized programs}
  $\mathit{worklist}\gets\{\operatorname{Initialize}(G,R)\}$\;
  $\mathit{seen}\gets\varnothing$\;
  \While{$\mathit{worklist}\neq\varnothing$}{
    $S\gets\operatorname{Take}(\mathit{worklist})$\;
    \If{$\exists T\in\mathit{seen}: S\approx_p T$
        \nllabel{ln:extract-equivalence}}{
      \textbf{continue}\nllabel{ln:extract-prune}\;
    }
    $\mathit{seen}\gets\mathit{seen}\cup\{S\}$\;
    \eIf{$S$ is complete}{
      output $\operatorname{Reconstruct}(S)$\nllabel{ln:extract-emit}\;
    }{
      choose an unresolved e-class $c$ in $S$\nllabel{ln:extract-choose}\;
      \ForEach{valid e-node $n\in c$}{
        $\mathit{worklist}\gets
          \mathit{worklist}\cup\{S[c\mapsto n]\}$
          \nllabel{ln:extract-extend}\;
      }
    }
  }
\end{algorithm}

Algorithm~\ref{alg:tensor-program-extraction} prunes redundant partial
programs before expansion and emits complete representatives for
performance evaluation.
Since $S\approx_p T$ implies that exploring $T$ covers every
candidate class reachable through $S$, pruning $S$ is sound and
preserves quotient-space coverage when enumeration completes.
A conservative test may miss redundant branches but does not
remove candidate classes.

\subsection{Searching Larger Graphs}
\label{sec:subgraph-composition}

For larger graphs, \system explores partitions with different subgraph
sizes. Larger subgraphs allow more joint transformations, while smaller
ones reduce search cost but may require more intermediate storage.
We rank partitions by estimated subgraph work and data transferred
across boundaries. Each subgraph is connected, and no dependency path
may leave it and later re-enter it.

We apply equality saturation and extraction to each distinct
subgraph, preserving its input/output interface. Candidates may
contain multiple kernels and are reused whenever the subgraph
appears in another partition. We combine them lazily through
their typed interfaces, storing intermediate results between
subgraphs without enumerating the full Cartesian product.

%% file: figures/extraction-quotient.tex
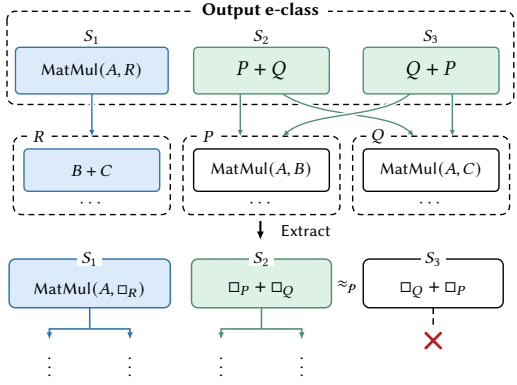
\begin{figure}[t]
  \centering
  \definecolor{compactionblue}{HTML}{DCEBFA}
  \definecolor{compactiongreen}{HTML}{DDF1E5}
  \resizebox{0.8\columnwidth}{!}{%
\begin{tikzpicture}[
  x=1pt,y=-1pt,
  label/.style={anchor=center,inner sep=0pt,
    font=\sffamily\fontsize{8}{10}\selectfont,text=black},
  enode/.style={rounded corners=2pt,line width=.65pt,
    minimum height=23pt,inner xsep=3pt,inner ysep=2pt,
    font=\fontsize{8}{10}\selectfont,text=black,fill=white},
  eclass/.style={rounded corners=4pt,dash pattern=on 2.8pt off 2pt,
    draw=black,line width=.65pt,fill=white},
  graph edge/.style={-{Latex[length=3pt,width=2.6pt]},
    line width=.65pt,draw=black,
    preaction={draw=white,line width=2.2pt}},
  search edge/.style={-{Latex[length=3.4pt,width=2.8pt]},
    line width=.75pt,draw=paperteal},
  omitted/.style={densely dashed,line width=.7pt,draw=black}
]
  \path[use as bounding box] (0,-5) rectangle (239,174);

  \filldraw[eclass] (0,0) rectangle (239,44);
  \node[label,fill=white,inner xsep=4pt,
    font=\sffamily\bfseries\fontsize{8.5}{10}\selectfont]
    at (119.5,0) {Output e-class};
  \node[label,text=black] at (40,12) {$S_1$};
  \node[label,text=black] at (120,12) {$S_2$};
  \node[label,text=black] at (200,12) {$S_3$};

  \draw[graph edge,draw=paperblue] (40,39) -- (40,59);
  \draw[graph edge,draw=paperteal] (110,39) -- (110,59);
  \draw[graph edge,draw=paperteal]
    (130,39) .. controls (148,51) and (180,47) .. (192,59);
  \draw[graph edge,draw=paperteal] (210,39) -- (210,59);
  \draw[graph edge,draw=paperteal]
    (190,39) .. controls (176,51) and (144,47) .. (130,59);

  \node[enode,minimum width=72pt,minimum height=22pt,
    draw=paperblue,fill=compactionblue]
    at (40,28) {$\mathsf{MatMul}(A,R)$};
  \node[enode,minimum width=64pt,minimum height=22pt,
    draw=paperteal,fill=compactiongreen,
    font=\fontsize{10}{12}\selectfont] at (120,28) {$P+Q$};
  \node[enode,minimum width=64pt,minimum height=22pt,
    draw=paperteal,fill=compactiongreen,
    font=\fontsize{10}{12}\selectfont] at (200,28) {$Q+P$};

  \filldraw[eclass] (3,59) rectangle (77,96);
  \node[label,fill=white,inner xsep=2pt,text=black] at (15,59) {$R$};
  \node[enode,minimum width=64pt,minimum height=20pt,
    draw=paperblue,fill=compactionblue] at (40,75) {$B+C$};
  \node[label,text=black] at (40,91) {$\cdots$};

  \filldraw[eclass] (83,59) rectangle (157,96);
  \node[label,fill=white,inner xsep=2pt,text=black] at (95,59) {$P$};
  \node[enode,minimum width=64pt,minimum height=20pt,
    draw=black] at (120,75) {$\mathsf{MatMul}(A,B)$};
  \node[label,text=black] at (120,91) {$\cdots$};

  \filldraw[eclass] (163,59) rectangle (237,96);
  \node[label,fill=white,inner xsep=2pt,text=black] at (175,59) {$Q$};
  \node[enode,minimum width=64pt,minimum height=20pt,
    draw=black] at (200,75) {$\mathsf{MatMul}(A,C)$};
  \node[label,text=black] at (200,91) {$\cdots$};

  \draw[-{Latex[length=4pt,width=3.5pt]},draw=black,line width=1pt]
    (119.5,99) -- (119.5,108);
  \node[label,anchor=west,font=\sffamily\fontsize{8}{10}\selectfont]
    at (129,103.5) {Extract};

  \filldraw[rounded corners=3pt,fill=compactionblue,
    draw=paperblue,line width=.7pt] (1,117) rectangle (77,140);
  \filldraw[rounded corners=3pt,fill=compactiongreen,
    draw=paperteal,line width=.7pt] (87,117) rectangle (153,140);
  \filldraw[rounded corners=3pt,fill=white,
    draw=black,line width=.65pt] (168,117) rectangle (234,140);
  \node[label,fill=white,inner xsep=2pt,text=black]
    at (39,117) {$S_1$};
  \node[label,fill=white,inner xsep=2pt,text=black]
    at (120,117) {$S_2$};
  \node[label,fill=white,inner xsep=2pt,text=black]
    at (201,117) {$S_3$};
  \node[label,font=\fontsize{8}{10}\selectfont]
    at (39,131.5) {$\mathsf{MatMul}(A,\square_R)$};
  \node[label,font=\fontsize{9}{11}\selectfont]
    at (120,131.5) {$\square_P+\square_Q$};
  \node[label,text=black,font=\fontsize{9}{11}\selectfont]
    at (201,131.5) {$\square_Q+\square_P$};
  \node[label,font=\fontsize{8}{10}\selectfont]
    at (160.5,130) {$\approx_p$};

  \draw[draw=paperblue,line width=.75pt] (39,140) -- (39,150);
  \draw[search edge,draw=paperblue] (39,150) -| (20,155);
  \draw[search edge,draw=paperblue] (39,150) -| (58,155);
  \node[label,text=black] at (20,165) {$\vdots$};
  \node[label,text=black] at (58,165) {$\vdots$};

  \draw[draw=paperteal,line width=.75pt] (120,140) -- (120,150);
  \draw[search edge] (120,150) -| (101,155);
  \draw[search edge] (120,150) -| (139,155);
  \node[label,text=black] at (101,165) {$\vdots$};
  \node[label,text=black] at (139,165) {$\vdots$};

  \draw[omitted] (201,140) -- (201,150);
  \draw[draw=paperred,line width=1.35pt,line cap=round]
    (197.5,152) -- (204.5,159) (197.5,159) -- (204.5,152);
\end{tikzpicture}
  }

  \caption{Early pruning of equivalent partial programs.}
  \Description{A shared output e-class contains a factored matrix
  multiplication and two sums with commuted operands. Shared child e-classes
  represent the input sum and the two matrix products. Extraction produces
  three partial programs with unresolved implementation choices. An equivalence
  marker relates the two commuted partial sums. The factored
  program and one sum continue expanding, while the duplicate sum branch is
  pruned before its remaining choices are explored.}
  \label{fig:extraction-quotient}
\end{figure}

%% file: sections/06-code-generation-and-tuning.tex
\section{Code Generation and Tuning}
\label{sec:code-generation-and-tuning}

Each extracted program and its legal schedule domain form an
implementation family. We canonicalize execution structures and
discard a family when another with the same structure covers its
entire schedule domain. For each remaining family, the backend
constructs a kernel graph, generates executable kernels under a legal
schedule, and measures their end-to-end latency.

\input{figures/diverse-tensor-program-results}

\subsection{Execution Families and Code Generation}
\label{sec:code-generation}

Code generation maps the IR's parallel structure to GPU execution.
Parallel axes that index output tiles determine the thread-block
grid, while parallel reduction axes are traversed by loops within
each block. The reducer body defines how the accumulated state is
updated, and operand views determine the tensor elements loaded
and stored. When intermediate results are materialized, producer
kernels write them to global memory for subsequent kernels to
consume. Kernel launches follow the program's data dependencies.

In M2 of Figure~\ref{fig:attention-rewrites}, the backend maps
axis $m$ to the thread-block grid and axis $p$ to a loop over
key/value tiles within each block. The fused reduction state
becomes loop-carried variables in the generated kernel.

For each candidate, the backend still needs to choose tile sizes,
the number of warps, pipeline stages, and implementations of matrix
multiplication and reductions. Once these choices are fixed, it
emits Triton~\cite{triton} kernels. Since different schedules can
yield different performance for the same candidate, comparing
candidates requires tuning and measuring their implementations.

\subsection{Static Family Ranking}
\label{sec:family-ranking}

Exhaustively tuning every candidate is expensive because each
has many possible schedules. We first rank candidates using
an optimistic estimate of the latency each could achieve under
a suitable schedule. This ranking determines which candidates
to profile first. Measurements then determine which candidates
to tune further.

For each candidate, we estimate computation and global-memory
traffic as functions of its schedule parameters. Dividing these
quantities by the corresponding hardware throughput or bandwidth
gives an optimistic execution time. We take the maximum across
resources to capture the bottleneck, then minimize over legal
schedules:
\begin{equation*}
  \hat{T}
  = \min_{\theta\in\mathcal{S}}
    \max_{r\in\mathcal{R}} \frac{W_r(\theta)}{P_r}.
\end{equation*}
Here $\mathcal{S}$ is the set of legal schedules, $\mathcal{R}$
is the set of modeled hardware resources, $W_r(\theta)$ is the
work or data movement for resource $r$, and $P_r$
is its throughput or bandwidth.

We use a linear programming relaxation to rank candidates without
enumerating every schedule. We then round the solution to obtain
an initial schedule for tuning.

\subsection{Progressive Schedule Tuning}
\label{sec:program-tuning}

We tune shortlisted candidates in rounds, using successive
halving~\cite{DBLP:conf/aistats/JamiesonT16} to give faster candidates
larger measurement budgets while keeping some candidates with low
estimated latency. For each candidate, we explore schedules
through local changes, evolutionary mutations, and random sampling.
For multi-kernel programs, we fix the parameters that determine
intermediate buffer shapes and tune one kernel at a time,
validating and timing the complete program at each step.
We return the fastest validated implementation found.

%% file: figures/diverse-tensor-program-results.tex
\begin{figure*}[t]
  \centering
  \includegraphics[width=\textwidth]{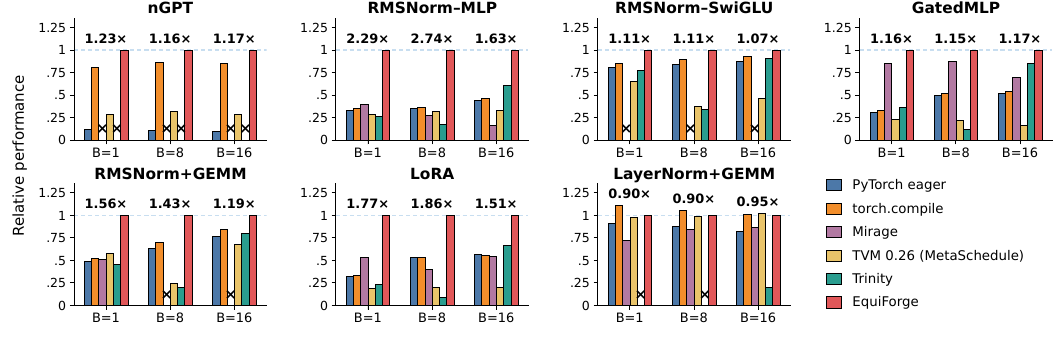}
  \caption{Performance across diverse tensor programs, normalized to \system.
    Higher is better. Labels show geometric mean speedups over the fastest valid baseline.
    $\times$: search timeout or numerically invalid result.}
  \Description{Seven small-multiple grouped bar charts compare PyTorch eager,
    torch.compile, Mirage when available, TVM 0.26, Trinity, and EquiForge
    across 21 non-attention configurations.  EquiForge is normalized to one.}
  \label{fig:diverse-tensor-programs}
\end{figure*}

%% file: sections/07-evaluation.tex
\section{Evaluation}
\label{sec:evaluation}

We evaluate \system's effectiveness and search efficiency on
tensor programs and Transformer layers.

\subsection{Experimental Setup}
\label{sec:experimental-setup}

\noindent\textbf{Implementation.}
Our frontend captures PyTorch~2.10 programs with \texttt{torch.export}~\cite{pytorch2}
and translates them into our IR. It expands composite operators such as
attention and RMSNorm into matrix multiplications, reductions, and pointwise
operations, allowing equality saturation to optimize their internal computations.
We use \textsc{egg}~0.11~\cite{egg} for saturation and extraction, and
Triton~3.6~\cite{triton} for code generation.

Our IR supports 18 pointwise operators plus reshape, transpose, and
tensor construction. We implement 120 rewrite rules:
21 algebraic, 10 refinement, 73 propagation, 9 reduction fusion, 6 structural,
and 1 for intermediate storage.

We use SymPy~\cite{sympy} to synthesize repair functions and check their
algebraic identities.
We use a four-hour budget per configuration for search and tuning.

\noindent\textbf{Workloads.}
Our workloads cover normalization, MLPs, low-rank adaptation,
and causal multi-head attention (MHA) in decode and prefill,
with batch sizes $B\in\{1,8,16\}$.
We use Mirage~\cite{mirage} and Prism~\cite{prism} benchmarks with
Llama~2~\cite{DBLP:journals/corr/abs-2307-09288} and BERT~\cite{bert}
configurations.
We explore various Transformer layers and highlight five cases:
QK-normalized multi-head latent
attention (MLA)~\cite{qknormmla}, mHC~\cite{mhc}, sliding-window
GQA~\cite{mistral7b}, attention sinks~\cite{streamingllm}, and differential
attention~\cite{differentialtransformer}.

\noindent\textbf{Baselines.}
For the non-attention workloads, we compare against PyTorch eager and
\texttt{torch.compile}~\cite{pytorch2}, the tensor compiler
TVM~\cite{tvm,tensorir}, and the tensor program superoptimizers
Mirage~\cite{mirage} and Trinity~\cite{trinity}.\footnote{We omit Prism because
its implementation was not publicly available at the time of evaluation.}
TVM receives 1,000 MetaSchedule
trials per shape. We use Trinity's default saturation and profiling budgets
and give Mirage the same four-hour search limit as \system. For
attention, we compare against the compilers Trinity and Neptune~\cite{neptune}
and the specialized FlashAttention~\cite{flashattention,flashattention2}
and FlashInfer~\cite{flashinfer} kernels.
Transformer-layer baselines use PyTorch eager execution and
\texttt{torch.compile}, with SDPA dispatching to FlashAttention.

\noindent\textbf{Hardware and measurement.}
All experiments run on NVIDIA A100 80GB PCIe and RTX 5090 GPUs.
We measure median tensor-program latency using CUDA Graphs, normalize each
system to \system on the same GPU, and report geometric means across GPUs.
Outputs are checked against PyTorch with a relative $L_2$ error tolerance of $0.01$.

\subsection{Tensor Program Performance}
\label{sec:program-performance}

\input{figures/attention-baseline-results}

\noindent\textbf{Diverse tensor programs.}
Figure~\ref{fig:diverse-tensor-programs} compares performance on the
tensor-program benchmarks. \system outperforms the fastest baseline in \ProgramWins{} of
\ProgramCases{} configurations, achieving a geometric mean speedup
of $\ProgramSpeedup\times$ and a maximum of $\ProgramMaxSpeedup\times$.

For example, on RMSNorm--MLP at $B=8$, \system moves the per-row normalization
factors $s$ after each projection using $(s\odot X)W \equiv s\odot(XW)$.
It then fuses both projections with the final elementwise product, avoiding
intermediate writes to global memory and achieving a $\PrismBEightCompileSpeedup\times$ speedup
over \texttt{torch.compile}.
The $B=1$ winner also applies normalization after both projections and
keeps their reductions in one kernel.

At $B=8$, Mirage also selects a single kernel for
RMSNorm--MLP, but normalizes the input before the projections.
Trinity uses three kernels, materializing the normalization
statistics and normalized activations.
LoRA at $B=1$ illustrates a different tradeoff. Trinity's
selected kernel recomputes the low-rank projection for each
output tile. \system uses two kernels, storing partial low-rank
products for reuse across output tiles. In our search,
this two-kernel candidate is $1.28\times$ faster than the
single-kernel candidate compiled with the same backend.

LayerNorm--GEMM remains the exception, running
$\LayerNormGapMin\%$--$\LayerNormGapMax\%$ slower than the fastest baseline.
The selected implementations use two, three, and four kernels at
$B=1,8,16$, respectively, computing normalization statistics before the
projection and incurring additional launches and intermediate memory traffic.

\input{figures/transformer-layer-structures}

\noindent\textbf{Attention.}
As shown in Figure~\ref{fig:attention-performance}, \system outperforms all compared baselines in the three decode cases. At $B=1$, the selected implementation
splits the output channels across two blocks per head. Each block computes
the same scores and softmax but produces a different part of the output.
This trades repeated computation for greater parallelism without requiring
a separate merge kernel.

In prefill, \system approaches FlashAttention's performance, with a latency
gap of $\MHAPrefillGapMin$--$\MHAPrefillGapMax\%$. The generated kernels combine
tiled matrix multiplications with an online softmax, incrementally updating
the normalization and attention output without storing the full score and
probability matrices. We use Triton for code generation, and its reference attention
kernel shows a similar gap to FlashAttention. This suggests that
the remaining gap may partly arise from scheduling and code
generation, including tile selection and the overlap of memory
accesses with computation.

\subsection{Transformer Layer Optimization}
\label{sec:layer-optimization}

We highlight five Transformer layers where \system discovers substantial
speedups. Each case uses one query token. Sizes below give
(batch size, hidden width, history length).

\noindent\textbf{QK-normalized MLA}~\cite{qknormmla}
$(1,4096,16\mathrm{K})$.
\system moves the key projection onto the query path and the value projection
after attention (Figure~\ref{fig:transformer-layer-case-studies}(a)).
This avoids expanding historical keys and values from the latent
cache, replacing projections over cached tokens with operations on a single
query and an aggregated context. Cached inverse-RMS factors move from the
expanded keys to the attention logits, preserving key normalization
in the transformed computation.
The resulting implementation achieves speedups of $3.53\times$ over
PyTorch eager and $3.16\times$ over \texttt{torch.compile}.

\noindent\textbf{mHC}~\cite{mhc}
$(16,512,1\mathrm{K})$.
The layer maintains four residual streams, with separate connections
around attention and the MLP.
As shown in Figure~\ref{fig:transformer-layer-case-studies}(b), \system
combines the first residual update with the next connection's normalization
in one kernel, allowing fusion across the model's component boundaries.
The normalization reuses the updated residual within the kernel,
avoiding an additional read of that intermediate.
The selected whole-layer implementation achieves
speedups of approximately $7.5\times$ over PyTorch eager and $5.84\times$
over \texttt{torch.compile}.

\noindent\textbf{Sliding-window GQA}~\cite{mistral7b}
$(1,4096,16\mathrm{K})$.
\system combines score computation, an online normalizer, and value
accumulation in one attention kernel. The normalizer processes only
the 4096-token window. Value accumulation still scans the full history
and recomputes score tiles, avoiding global score and probability
matrices at the cost of additional arithmetic.
This implementation achieves speedups of $1.12\times$ over
PyTorch eager and $1.06\times$ over \texttt{torch.compile}.

\noindent\textbf{Attention sinks}~\cite{streamingllm}
$(1,1024,32\mathrm{K})$.
\system fuses score computation, online normalization, and value accumulation
into one attention kernel. It visits only the tiles covering four initial
sink tokens and the most recent 4096 tokens, skipping the masked middle
without materializing score or probability matrices.
Both ranges contribute to one online reduction, so the sink and recent
tokens share the same softmax normalization.
The seven-kernel layer runs $2.18\times$ faster than PyTorch eager
and $1.48\times$ faster than \texttt{torch.compile}.

\noindent\textbf{Differential attention}~\cite{differentialtransformer}
$(1,1024,32\mathrm{K})$.
The two attention branches share a value tensor. Their learned
weighted difference feeds per-head RMSNorm.
\system evaluates the mixing weight, subtracts the attention outputs,
and normalizes each head in one kernel. The mixed output is consumed
directly by normalization, avoiding an intermediate tensor and separate
launches for these operations.
This implementation achieves speedups of $1.66\times$ over
PyTorch eager and $1.51\times$ over \texttt{torch.compile}.

\subsection{Search Cost}
\label{sec:search-cost}

\textbf{Search overhead.}
Table~\ref{tab:search-space-size} reports e-graph sizes and the time
spent in saturation and extraction at $B=16$. Most non-attention workloads produce
fewer than 1,200 e-nodes and complete these stages within tens of
seconds. Attention produces much larger e-graphs, with over 22,000
e-nodes in decode and 65,000 in prefill. Its two matrix
multiplications and dependent softmax reductions expose interacting
choices in tiling, fusion, and intermediate storage. Combining these
choices creates many equivalent implementations, increasing both
saturation and extraction costs. These results highlight the cost
of exploring transformations jointly and motivate early compaction
to avoid repeatedly constructing redundant candidates.

\input{figures/search-space-size-table}


\textbf{Extraction compaction.}
We compare three strategies.
\begin{itemize}[leftmargin=0.5\leftmargini, itemsep=1pt]
  \item \textbf{No compaction} enumerates every complete candidate.
  \item \textbf{Compaction after completion} constructs all candidates and then
    selects one per equivalence class.
  \item \textbf{Early compaction} prunes equivalent partial programs before
    completing them.
\end{itemize}

\input{figures/extraction-compaction}

Figure~\ref{fig:extraction-compaction} reports candidate counts
and enumeration time. Compaction after completion and early
compaction use the same e-graph and compaction equivalence
relation. Across the four workloads where both strategies
finish, early compaction reduces the number of expanded states
by factors of $128$--$2{,}465$ while returning the same number
of distinct candidates. For nGPT, it reduces enumeration time
from $106$\,s to $0.47$\,s. For RMSNorm--SwiGLU, early compaction
finishes with 330 candidates in $16.97$\,s, whereas compaction
after completion times out after two hours with only 72.
This shows that early compaction reduces extraction
cost and can return more distinct candidates within a fixed
time budget.

%% file: figures/attention-baseline-results.tex
\begin{figure}[t]
  \centering
  \includegraphics[width=\columnwidth]{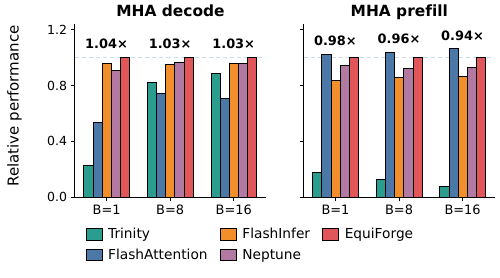}
  \caption{MHA performance in decode and prefill, normalized to \system.
    Higher is better. Labels show geometric mean speedups over the fastest valid baseline.}
  \Description{Two side-by-side grouped bar charts compare Trinity,
    FlashAttention, FlashInfer, Neptune, and EquiForge for MHA decode and prefill
    at batch sizes 1, 8, and 16.  EquiForge is normalized to one.}
  \label{fig:attention-performance}
\end{figure}

%% file: figures/transformer-layer-structures.tex
\begin{figure*}[t]
  \centering
  \includegraphics[width=\textwidth]{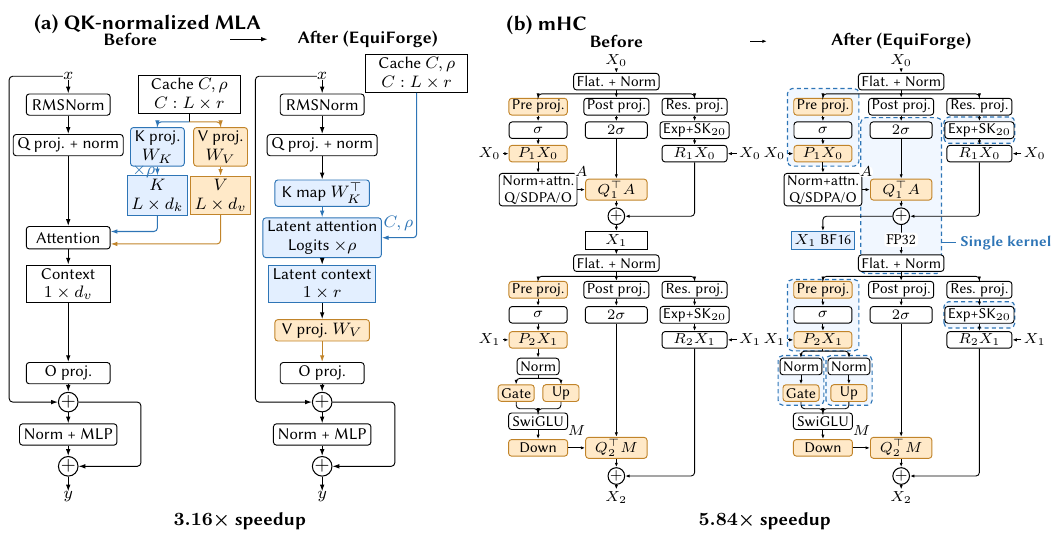}
  \caption{Layer structures found by \system: (a) QK-normalized MLA and (b) mHC.
    Each dashed blue box encloses one kernel. SK$_{20}$ denotes 20 Sinkhorn iterations.
    Panel (a) shows logical dataflow: cached inverse-RMS key statistics $\rho$
    scale keys before rewriting and logits afterward; auxiliary state outputs
    and tiled partial reductions are omitted.}
  \Description{Two side-by-side case studies each compare source and optimized layer
    dataflow. Labels report whole-layer speedups over torch.compile, whose SDPA
    backend uses FlashAttention: 3.16 times for QK-normalized MLA and 5.84 times
    for mHC. In QK-normalized MLA, the key projection moves to the query path
    and the value projection follows attention, eliminating expanded
    historical key and value tensors. Both MLA views retain the cached inverse-RMS
    key statistics rho: they scale projected keys before the rewrite and
    attention logits afterward. The MLA schematic omits the auxiliary new-latent
    and inverse-RMS outputs, tiling, and partial reductions. The full mHC block has independent
    pre, post, and residual maps around attention and the MLP. Seven dashed
    blue boxes show two fused pre-mix chains, the first post gate through
    residual addition and the next map normalization, two separate
    normalization-plus-projection kernels for the MLP, and two fused Sinkhorn
    computations. A blue Single kernel label points to the box spanning the
    two connections. The first residual is returned in BF16 while its FP32 value
    feeds the next map normalization.}
  \label{fig:transformer-layer-case-studies}
\end{figure*}

%% file: figures/search-space-size-table.tex
\begin{table}[htbp]
  \caption{\system e-graph size and search cost at $B=16$.}
  \label{tab:search-space-size}
  \centering
  \small
  \begin{tabular}{@{}lrrr@{}}
    \toprule
    & & \multicolumn{2}{c}{Time (s)} \\
    \cmidrule(lr){3-4}
    Workload & E-classes/E-nodes & Sat. & Extract. \\
    \midrule
    nGPT            & 100/296      & 0.09   & 1.30  \\
    RMSNorm--MLP    & 346/949      & 8.72   & 4.34  \\
    RMSNorm--SwiGLU & 411/1,179    & 0.89   & 21.37 \\
    GatedMLP        & 256/645      & 0.04   & 0.91  \\
    RMSNorm+GEMM    & 204/590      & 0.02   & 0.51  \\
    LoRA            & 284/870      & 0.07   & 4.06  \\
    LayerNorm+GEMM  & 1,057/5,388  & 151.6  & 117.4 \\
    \midrule
    MHA decode      & 5,223/22,339 & 395.3  & 272.4 \\
    MHA prefill     & 5,286/65,283 & 1741.9 & 910.7 \\
    \bottomrule
  \end{tabular}
\end{table}

%% file: figures/extraction-compaction.tex
\begin{figure}[t]
  \centering
  \includegraphics[width=\columnwidth]{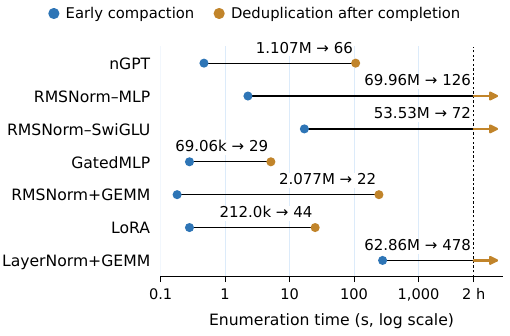}
  \caption{Enumeration time and candidate counts at $B=16$.  Labels show
    complete candidates constructed $\to$ distinct candidates after
    deduplication.}
  \Description{A horizontal paired-dot chart compares early compaction in
    blue with deduplication after completion in orange for seven workloads.
    Enumeration time is plotted on a logarithmic axis.  Numeric labels above
    the connectors report complete candidate counts before and after
    deduplication. Orange arrows indicate
    unfinished runs for RMSNorm--MLP, RMSNorm--SwiGLU, and LayerNorm+GEMM.}
  \label{fig:extraction-compaction}
\end{figure}

%% file: sections/08-related-work.tex
\section{Related Work}
\label{sec:related-work}

We review prior work on tensor program optimization and equality saturation.

\noindent\textbf{Tensor compilers and optimization.}
Halide~\cite{halide} separates algorithms from schedules.
TVM~\cite{tvm} combines graph optimization with automated operator code
generation, while TensorIR~\cite{tensorir} integrates tensorization into
scheduling. Ansor~\cite{ansor} uses evolutionary schedule
search and a learned cost model. ROLLER~\cite{DBLP:conf/osdi/ZhuWDKLZXMXC0YZ22}
constructs hardware-aware tiles, while
Felix~\cite{DBLP:conf/asplos/ZhaoSAM24} uses continuous relaxation and a
differentiable latency model to guide search.
\system complements schedule optimization with joint search
over algebraic forms and parallel decompositions.

\noindent\textbf{Tensor program superoptimization.}
TASO~\cite{taso} searches tensor graphs using automatically generated and
verified substitutions. PET~\cite{pet} permits partially equivalent
transformations with generated corrections that restore full equivalence.
TENSAT~\cite{tensat} and SPORES~\cite{DBLP:journals/pvldb/WangHSHL20}
optimize tensor expressions through equality saturation.
Unity~\cite{unity} jointly optimizes graph rewrites and distributed
parallelization strategies.

Mirage~\cite{mirage} and Prism~\cite{prism} are enumeration-based
superoptimizers. Mirage enumerates concrete programs across kernel,
thread-block, and thread levels and uses probabilistic verification.
Prism enumerates symbolic program families and verifies them through
e-graph rewriting. \system generates equivalent implementations
through equality saturation and applies early compaction during extraction.

Trinity~\cite{trinity} applies
equality saturation to a tile-level IR with explicit loops and memory
operations. This representation requires expression propagation to expose
algebraic rewrites across loads and stores, and sequence canonicalization to
control e-graph growth. Its two-pass extractor selects loop structures before their
bodies. \system uses a pure expression IR in which tensor-level
algebraic rewrites compose with parallel refinement.

\noindent\textbf{Operator fusion.}
Welder~\cite{DBLP:conf/osdi/00010XMXMG0Z23} balances data reuse with tile
graphs, while Korch~\cite{korch} decomposes operators to optimize kernel
orchestration. Neptune~\cite{neptune} derives algebraic repairs for fusion
of dependent reductions, and Nautilus~\cite{nautilus} combines these
transformations with automatic scheduling. Flashlight~\cite{flashlight}
generates fused kernels for attention variants through PyTorch compiler
extensions. \system expresses Neptune's repair relation as an equality
that composes with parallel refinement and tensor rewrites.

\noindent\textbf{Equality saturation and extraction.}
Tate et~al.~\cite{DBLP:conf/popl/TateSTL09} introduced equality saturation,
and \textsc{egg}~\cite{egg} supports efficient rewriting through rebuilding
and e-class analyses.
Guided equality saturation~\cite{DBLP:journals/pacmpl/KoehlerGBGTS24} uses
program sketches to direct rewriting. Recent algorithms exploit e-graph
structure to accelerate optimal
extraction~\cite{DBLP:journals/pacmpl/GoharshadyLP24,DBLP:journals/corr/abs-2408-17042}.
Equality-constrained tree automata~\cite{DBLP:journals/pacmpl/KoppelGVSP22}
support enumeration under shared constraints. \system enumerates candidates
for performance measurement and prunes redundant partial programs.

%% file: sections/09-conclusion.tex
\section{Conclusion}
\label{sec:conclusion}

We presented \system, a tensor program superoptimizer that brings algebraic
reformulation and parallel implementation generation into the same equality
search.  Its unified typed IR allows these transformations to compose
throughout the search, while compaction prunes equivalent branches before
candidate programs are complete. Composing equality rules yields fused
implementations such as online attention directly from tensor expressions.
Subgraph composition extends the search to larger programs and enables
optimization across Transformer-layer components. Our evaluation shows a
$\ProgramSpeedup\times$ geometric mean speedup over the fastest baseline for
each tensor-program configuration. Early compaction cuts the number of states
expanded during extraction by a factor of up to $2{,}465$.
These results support a design in which
equality reasoning determines program structure and controls redundant
exploration.